\def\gp{\gamma_+}
\def\gm{\gamma_-}

\def\gt{\gamma_t}
\def\gcm{\gamma_{cm}}
\def\gc{\gamma_c}
\def\gr{\gamma_r}
\def\bcm{\beta_{cm}}
\def\bc{\beta_c}
\def\br{\beta_r}
\def\bt{\beta_t}

\def\game{\gamma_1}
\def\gz{\gamma_2}

\def\ugr{\> \lower4pt \hbox{$\buildrel > \over \sim$} \> }
\def\ukl{\> \lower4pt \hbox{$\buildrel < \over \sim$} \> }

\def\bcmv{\overrightarrow\bcm}
\def\bcv{\overrightarrow\bc}

\def\bg{\beta_{\Gamma}}

\def\be{\beta_1}
\def\bz{\beta_2}
\def\Npm{N_{\pm}}

\def\gpm{\gamma_{\pm}}
\def\gmp{\gamma_{\mp}}
\def\bpm{\beta_{\pm}}

\def\gepm{\gamma_{1\pm}}

\def\gzpm{\gamma_{2\pm}}

\def\de{\Delta \gamma}

\def\deqm{\left\langle (\de)^2 \right\rangle}

\def\kq{\overline K (\gcm)}
\def\st{\sigma_T}
\def\dt#1{{\partial #1 \over \partial t}}
\def\gamr{gamma-ray }
\def\lnl{\ln\Lambda}
\def\ecm{\epsilon_{cm}}

\documentstyle[11pt, aaspp4, epsf, rotate]{article}

\lefthead{B\"ottcher, Pohl \& Schlickeiser}
\righthead{Transrelativistic pair plasmas in AGN jets}

\begin{document}

\title{Transrelativistic pair plasmas in AGN jets}

\author{M.  B\"ottcher \altaffilmark{1}, M. Pohl \altaffilmark{2},
R. Schlickeiser \altaffilmark{3}}

\altaffiltext{1}{Rice University, MS 108, Space Physics and Astronomy 
Department, \\ 6100 S. Main Street, Houston, TX 77005 -- 1892, USA; 
E-mail: mboett@spacsun.rice.edu \\ phone: (713) 527 8750 Ext. 2653,
fax: (713) 285 5143}

\altaffiltext{2}{Danish Space Research Institute, Juliane Maries Vej 30,
DK -- 2100 K\o benhavn \O, Denmark}

\altaffiltext{3}{Institut f\"ur Theoretische Physik, Lehrstuhl IV,
Ruhr-Universit\"at Bochum, \\ D -- 44 780 Bochum, Germany}

\begin{abstract}
Models of relativistic jets filled with ultrarelativistic
pair plasma are very successful in explaining the broadband 
radiation of $\gamma$-ray blazars. Assuming that the initial
injection and cooling of ultrarelativistic pair plasma in an AGN
jet has occurred, producing the observed high-energy $\gamma$-ray 
radiation, we investigate the further evolution of the pair
plasma as it continues to move out from the central engine. 
The effects of thermalization and reacceleration, the emission 
of pair bremsstrahlung and annihilation radiation and the bulk
Compton process, and the possible application to MeV blazars are 
discussed. A model calculation to the special case of PKS~0208-512 
is presented.
\end{abstract}

\keywords{acceleration of particles --- plasmas --- 
radiation mechanisms: non-thermal --- radiation mechanisms: thermal --- 
radiative transfer --- galaxies: jets --- gamma-rays: theory}

\section{Introduction}

The detection of high-energy $\gamma$-ray emission from more 
than 60 blazars with EGRET is a challenge and at the same time 
a constraint of fundamental importance for emission models
(von Montigny et al. 1995). A large fraction of these blazars
exhibits variability at $\gamma$-ray energies on time scales 
of days to months (Mukherjee et al. 1997). The optical 
counterparts of the majority of EGRET detected AGN are
known as BL Lacertae objects and optically violent variable quasars (OVV).
At radio wavelengths, all blazars can be recognized as bright, compact
sources with a flat synchrotron spectrum emanating from outflowing plasma
jets that are nearly aligned with our line-of-sight. Relativistic
beaming is required in the objects in view of the luminosity and variability
time scales (Dermer and Gehrels 1995), in accord with VLBI observations
indicating that superluminal motion is a common feature in this class of AGN
(e.g. Wehrle et al. 1994, Pohl et al. 1995, Barthel et al. 1995, Krichbaum
et al. 1995).

The strongest EGRET blazar detections can be characterized by a single 
power-law spectrum with differential photon spectral indices between 
$\alpha = 1.5$ and $\alpha = 2.7$ (Thompson et al. 1995). For 
individual sources the spectral index is correlated with the 
flux level, and there may also be deviations from the power-law 
behaviour both below 70 MeV and above a few GeV (Pohl et al.
1997). The combined OSSE/COMPTEL/EGRET measurements generally indicate
spectral breaks at a few MeV (Williams et al. 1995, McNaron-Brown et al. 
1995) in the sense that the spectra below 1 MeV are harder than in
the EGRET range.

At medium $\gamma$-ray energies observable by COMPTEL, PKS 0208-512
has been identified as an AGN with flaring properties at MeV energies
(Blom et al. 1995). There is now evidence that it may belong to a
class of `MeV-blazars' that are occasionally exceptionally bright
MeV sources (see Bloemen et al. 1995). Since in these objects the 
bright emission appears to be confined to a relatively narrow energy 
range, the discussion has focused on models involving a broad 
blue-shifted $e^+/e^-$ annihilation line that is Doppler boosted 
in a relativistic jet (Roland and Hermsen 1995). If this 
interpretation applies, MeV-blazars provide a unique tool to 
study astrophysical particle beams (e.g. Schlickeiser 1996).
Furthermore, MeV-blazars may confuse the analysis of galactic 
sources of $\gamma$-ray line emission (Pohl 1996).

In an earlier paper (B\"ottcher, Mause \& Schlickeiser 1997; 
hereafter BMS) we have investigated the temporal evolution 
of ultrarelativistic pair plasmas in jets of quasars and 
BL-Lac objects and have demonstrated that their
broadband spectra can well be explained as the resulting 
synchrotron and inverse-Compton radiation from a cooling
ultrarelativistic nonthermal pair plasma in a relativistic jet. 
A decisive parameter for the evolution of single ultrarelativistic
plasma components inside an AGN jet is the density of pairs injected 
into the jet. Broadband fits, covering the radio to $\gamma$-ray
regime of the electromagnetic spectrum, to blazars generally require 
particle densities of order $n_e \ukl 10^3$~cm$^{-3}$. Nevertheless, 
fits to different objects suggest that the value of the pair density
in relativistic jets ejected by active galactic nuclei varies over 
several orders of magnitude. Jets of very high density ($n \ugr
10^{5}$~cm$^{-3}$) can also produce the broadband spectra at least
of $\gamma$-ray active flat-spectrum radio quasars whose bolometric
luminosity is clearly dominated by the $\gamma$-ray emission.

In this paper, we argue that after an initial phase of rapid
cooling, governed by synchrotron and inverse-Compton energy
losses, the relativistic pair plasma inside such components is 
likely to attain a quasi-thermal distribution. If the jet
remains well collimated and cooling (e. g. through adiabatic
losses) remains very efficient through the transrelativistic
phase, pair bremsstrahlung and pair annihilation become
efficient. The resulting radiation spectrum peaks around
several MeV, which has been suggested previously to be
responsible for the observed MeV bump in MeV blazars
(Henri et al. 1993, Roland \& Hermsen 1995, B\"ottcher
\& Schlickeiser 1996). However, while generally the high-energy 
radiation from a few ultrarelativistic jet components, ejected 
over the typical EGRET variability time scale, is 
appropriate to model EGRET spectra, the pair annihilation
and bremsstrahlung radiation from only a few components after
cooling and thermalization is usually too weak to explain 
the MeV blazar phenomenon. Therefore, a quasi-continuous
supply of mildly relativistic pair plasma into the jet
is required in order to fit the MeV bump in MeV blazars
with pair annihilation radiation. 

Alternatively, if very close to the central accretion disk 
the inverse-Compton cooling rate is balanced by reacceleration
by hydromagnetic turbulences the pair plasma in the jet will
thermalize at relativistic temperatures and will continuously 
Compton upscatter external radiation from the accretion disk 
very efficiently, producing another bump at keV --- MeV energies 
by the so-called bulk Compton process (Sikora et al. 1997).

We know from the EGRET data of many blazars that the GeV emission 
is not that of a single injection of particles, but is more likely
to result from more or less regularly repeating injection events
with varying energy input into the ejected particles. A second 
important point is the correlation between flux and spectral index 
in the EGRET range. The spectral hardening during outbursts indicates 
a more efficient acceleration of particles. Even for highly variable
sources like PKS~0208-512, PKS~0528+134 (Collmar et al. 1997) and
others the low-energy $\gamma$-ray continuum varies with small 
amplitudes and not in phase with the variability in the EGRET range.
To be more realistic we may thus either assume that injections occur 
in regular time intervals, so that they would not influence each other 
in their evolution, or that injection occurs in a quasi-steady manner, 
such that relativistic pair plasma is continuously injected. 

For ease of both computing and exposition we will not 
consider the effect of the finite light travel time within 
the volume occupied by pair plasma, which limits our 
predictions of variability to time scales longer than the light
travel time through individual plasma blobs. In section 2, we 
give a short overview of the different elementary processes 
which play a role for the evolution of a relativistic pair 
plasma component of an AGN jet after the initial phase of 
rapid cooling and which are usually not considered in 
models of ultrarelativistic jets. In section 3, we 
describe numerical simulations and an analytical 
approximation, applicable under special conditions, 
to follow the evolution of the pair plasma through 
the transrelativistic phase. We find that, depending
basically on the particle density and the magnetic field, 
there are two different ways how a quasi-thermal distribution 
can be established. The relevant physics of these quasi-thermal 
plasmas is described in section 4. Section 5 contains a model 
calculation for the typical MeV blazar PKS~0208-512. We summarize 
in section 6.

\section{The elementary processes}

The momentum distribution of pairs inside a relativistic
jet component is assumed to be isotropic and (for sake
of simplicity) homogeneous. Under circumstances
which will be specified below, the time evolution 
of such a pair plasma can be approached by a Fokker-Planck 
equation regarding only time and particle energy as variables:

\begin{equation}
{\partial \over \partial t} n (\gamma, t) + {\partial \over
\partial \gamma} \left[ \left( {d \gamma \over d t} \right)
n (\gamma, t) \right] 
- {1 \over 2} {\partial^2 \over \partial \gamma^2} \, \left[
{d \left(\Delta\gamma\right)^2 \over d t} \, n (\gamma, t) \right]
\> = \> {d \, n (\gamma, t) \over d t }.
\end{equation}
Here, ${d \gamma \over d t}$ is the single-particle energy loss 
due to all the relevant elementary processes and 
${d \left(\Delta\gamma\right)^2 \over d t}$ is the energy dispersion 
rate. The term on the right-hand sight of Eq. (1) is the source 
function. We restrict ourselves to the case where the plasma
component is optically thin to $\gamma$-$\gamma$ pair production
(this process might well have led to re-injection of pairs
during the ultrarelativistic phase but becomes irrelevant in 
the transrelativistic regime), and consider the catastrophic 
pair annihilation losses and the dilution of the blob due 
to expansion. 

A detailed description of the calculation of the energy loss/gain 
rates due to inverse-Compton scattering of accretion disk radiation, 
synchrotron emission and the SSC process can be found in BMS. At the 
final states of the simulations carried out there, the particles have 
cooled down to energies such that all Compton scattering events 
may be treated in the Thomson regime. This guarantees that the use 
of the Fokker-Planck equation to describe the effect of inverse-Compton
scattering is a good approach, which requires that the change in
particle energy is $\Delta\gamma \ll \gamma - 1$ for 
each scattering event. Since $\Delta\gamma \ukl \epsilon \, 
(\gamma^2 - 1) = {\epsilon (\gamma + 1)} \, {(\gamma - 1)}$ and
$\epsilon\gamma \ll 1$, this condition is fulfilled in the case of
Thomson scattering. The external photon source is assumed to
be dominated by the accretion disk. The full angle-dependence 
of the accretion-disk photon field is included in our calculation.

Modelling of blazar spectra indicates that the magnetic fields
in many AGN jets can be significantly lower than the equipartition 
value. Therefore, the gas pressure generally exceeds the magnetic
pressure, leading to a conical jet and freely expanding blobs. 
The inferred low magnetic fields (typically $B \ukl 0.1$ -- $1$~G) 
in some objects, together with the fact that the magnetic field
generally declines outward, imply that synchrotron radiation 
(and therefore also reacceleration by synchrotron-self absorption) 
is of minor importance in the evolutionary phase with which we are 
dealing in this paper.

In ultrarelativistic jets, the effects of elastic (M\o ller and
Bhabha) and inelastic scattering (pair bremsstrahlung emission),
adiabatic losses, pair annihilation and stochastic reacceleration 
may be neglected compared to the dominant influence of synchrotron 
and inverse-Compton losses (BMS). Being interested in the details 
of the process of cooling down to mildly-relativistic energies, 
we now have to consider them and therefore discuss their influence 
in the following subsections. In this phase of the jet evolution, 
also the expansion of the jet will have a significant influence 
via adiabatic cooling and the dilution of the particle densities, 
affecting elastic and inelastic scattering and pair annihilation.

\subsection{Elastic scattering}

The energy loss/gain rate due to M\o ller scattering of 
electrons (positrons) of energy $\gamma_t$ off an isotropic
distribution $n_{\pm} (\gpm) = 4\pi N_{\pm} \bpm\gpm^2 \, 
f_{\pm} (\gpm)$ of positrons (electrons) in the general case
has been calculated by Dermer (1985). Here, we may restrict
our considerations to the relativistic case in which the
elementary single-particle energy loss is dominated by
the term involving the Coulomb logarithm. 

As was suggested by Nayakshin \& Melia (1998), in this
case the single-particle energy gain/loss rate reduces
to a one-dimensional integral:

\begin{equation}
{d\gamma_t \over dt} = \int\limits_1^{\infty} d\gz \> n_{\pm} (\gz) \,
a(\gamma_t, \gz)
\end{equation}
where

\begin{equation}
 a(\gamma_t, \gz) = (\gz - \gamma_t) \, {2 \, \pi \, c \, r_e^2
\> A \over \gamma_t^2 \bt \, \gz^2 \bz} \> \left[ \gr \br
- {2 \over \br} + {\rm arcosh} \gr \right]_{\gr = \gamma_t \gz
(1 - \bt\bz)}^{\gamma_t \gz (1 + \bt\bz)}
\end{equation}
is the monoenergetic energy exchange rate. Here,

\begin{equation}
A = \ln\Lambda + {1 \over 2} \ln(2 e) \approx \ln\Lambda + 0.8466.
\end{equation}
The Coulomb logarithm $\ln\Lambda$ is only weakly dependent on the 
electron and positron energies. In our numerical calculations, we 
use the constant value $A = 20$. The difference between M\o ller 
(electron-electron) and Bhabha (electron-positron) scattering 
only occurs in the terms neglected in the derivation of eq. (3) 
and is negligible in the case of relativistic pairs. 
The energy-exchange rate scales linearly with particle density
and, for energies smaller than the average particle energy of the
distribution, $\gamma < \langle\gamma\rangle$, it declines roughly
as $d\gamma / dt \, \propto \gamma^{-1}$. It changes the sign at
$\gamma = \langle\gamma\rangle$ and approaches a constant, negative 
value for $\gamma \gg \langle\gamma\rangle$.

In order to evaluate the energy dispersion due to elastic (M\o ller
and Bhabha) scattering, we use the technique outlined by Dermer (1985),
together with the idea of Nayakshin \& Melia (1998) to consider
scattering off monoenergetic distributions (for which the dispersion
rate can be calculated analytically) and then averaging over the
distributions of the background particles. In contrast to the
technique of Dermer (1985), this requires only one numerical 
integration. The resulting expressions are given in Appendix A.1.
The energy-dispersion rate also scales linearly with particle density 
and depends only weakly on the particle energy in the range 
$\vert \gamma - \langle\gamma\rangle \vert \ll \langle\gamma\rangle$.
The Fokker-Planck equation (1) is only valid if the energy
distribution is narrow in the above sense, because otherwise
elastic scattering events between particles of very high
and very low energies would lead to a considerable energy
gain $\Delta\gamma \ugr \gamma$ of the low-energetic particle.

\subsection{Bremsstrahlung emission}

The general treatment of the pair bremsstrahlung process is very
cumbersome and has been investigated in several papers by Haug,
posing special interest on pair bremsstrahlung emission 
in thermal pair plasmas (Haug 1975, 1985a, 1985b). The
importance of this radiation mechanism to the hard X-ray and
$\gamma$-ray emission of relativistic jets in AGNs has been
demonstrated by B\"ottcher \& Schlickeiser (1995).

For the pair distributions considered in this paper, we may 
restrict ourselves to the ultrarelativistic limit using the 
differential pair bremsstrahlung cross section derived by 
Ba\u\i er, Fadin \& Khoze (1967). The bremsstrahlung 
energy-loss rate is calculated in Appendix A.2. 
Inspecting the function $\overline K$ given there,
we find that the final-state averaged energy-loss
of a test particle emitting brems\-strah\-lung photons is
typically of the order $m_e c^2$ and thus, for relativistic 
particles the necessary condition $\Delta\gamma \ll 
\gamma - 1$ is fulfilled. The energy-loss rate has the 
same linear dependence on particle density as for elastic 
scattering and depends on particle energy as $- d\gamma / dt 
\propto \gamma^{1.1}$. It is only weakly dependent on the 
shape of the relativistic pair distribution.

\subsection{Effects of jet expansion}

For the parameter range under consideration in this
paper, we expect the jet to expand freely. This yields 
a declining particle density as

\begin{equation}
n (t) = n_0 \, \left( {z (t) \over z_0} \right)^{-2}
\end{equation}
and
\begin{equation}
\left( {d n (t) \over dt} \right)_{\rm exp} =
- 2 \, {n (t) \over z} \, c \, \bg \, \Gamma.
\end{equation}
The expansion also leads to adiabatic losses due to the
invariance of the magnetic flux through the particle orbits,
$B \, r_L^2 =$~const., where $r_L$ is the Larmor radius of a
particle. This provides an additional energy-loss term,
$\dot\gamma = \gamma \beta^2 \, \dot B / (2 \, B)$, depending 
on the $z$ dependence of the magnetic field. In the case of a
perfectly isotropized magnetic field, its evolution is given
by $B(z) \propto z^{-2}$, while a purely transversal magnetic
field evolves as $B(z) \propto z^{-1}$ (Blandford \& Rees
1974, Blandford \& K\"onigl 1979). Since the degree of
isotropization of the magnetic field is not well known,
we parametrize the magnetic field dependence as

\begin{equation}
B(z[t]) = B_0 \, \left( {z(t) \over z_0} \right)^{-b}
\end{equation}
with $1 \le b \le 2$. This yields

\begin{equation}
\left( {d\gamma \over dt} \right)_{\rm ad} = - {b \over 2}
\, {\gamma \beta^2 \over z} \, \bg \, \Gamma \, c.
\end{equation}

\subsection{Stochastic acceleration by Alfv\`en waves}

In order to estimate the reacceleration and energy
dispersion rate due to wave-particle interaction we will
restrict our calculation to parallel, transverse plas\-ma waves,
i. e. ${\bf k} \parallel {\bf B}_0$ and ${\bf E} \bot 
{\bf B}_0$, in a pure pair plasma. Here, ${\bf k}$ is the wave
vector, ${\bf E}$ is the polarization vector of the plasma 
wave and ${\bf B}_0$ is the background magnetic field. 

According to the scenario of the formation of jet components, 
described in detail in BMS, the role of protons inside 
the jet may be neglected. In this scenario, the pairs are 
created as secondary particles in photo-pair and photo-pion 
production by relativistic protons accelerated in the 
accretion disk magnetosphere and will greatly outnumber the
primary particles and carry the dominant fraction of the
total energy transferred to particles, if the duty cycle of 
$\gamma$-ray emission is much less than unity. This pair population
becomes unstable with respect to the excitation of various
electromagnetic and electrostatic waves, resulting in an 
explosive event which ejects the ultrarelativistic pair 
plasma along an existing jet structure. This implies that
the mechanism for acceleration of the protons during the
quiescent phase by stochastic processes, ultimately
leading to the production of the unstable pair plasma
population, is of completely different nature than the 
mechanism which injects the pair plasma into the jet.
The magnetic field and the level of turbulence during
the latter phase, in turn, are in general drastically 
different from their values along the jet structure.

It is well-known that relativistic particles can only 
interact efficiently with the Alfv\`en part of the plasma 
wave spectrum. For this reason, we neglect interaction 
with electron (positron) cyclotron waves. The relevant 
Fokker-Planck coefficients for the interaction of
electrons with Alfv\`en wave turbulences propagating through an
electron-proton plasma have been investigated in detail by
Schlickeiser (1989). Here, we generalize his calculation to
the case of Alfv\`en waves in a pair plasma. We assume that 
the Alfv\`en waves have wave numbers $k$ between

\begin{equation}
k_{min} = {2 \, \pi \over R_B} \hskip 0.8cm {\rm and} \hskip 0.8cm
k_{max} = {\Omega_0 \over v_a}
\end{equation}
where $R_B$ is the radius of the jet component, $\Omega_0$
is the nonrelativistic gyrofrequency of electrons/positrons, and
$v_a = B_0 \sqrt{\langle\gamma\rangle} / \sqrt{8 \pi m_e n_e}$ 
is the Alfv\`en velocity in a relativistic pair plasma.

Eq. (9) certainly underestimates the minimum wave number since
we assume the magnetic field to be ordered on smaller scales than 
$R_B$. It is especially the highest-energetic particles which 
interact with these longest waves. But for ultrarelativistic particles, 
synchrotron and inverse-Compton cooling is much more important than
wave-particle interactions. Thus, the inaccuracy of Eq. (9) does 
not influence our final results. We assume that the spectral energy 
content in plasma waves is distributed according to a power-law,

\begin{equation}
I (k) = I_0 \, k^{-q}
\end{equation}
with
\begin{equation}
I_0 = 4 \, \pi \, \left( {\delta B \over B_0} \right)^2 {B_0^2 \over
8 \, \pi} {q - 1 \over k_{min}^{1 - q} - k_{max}^{1 - q}}.
\end{equation}
Here, $\delta B$ is the amplitude of the magnetic field fluctuation
due to the plasma waves which is typically of order $\ukl 10^{-1} \> 
B_0$. Now, following the calculations of Schlickeiser (1989) (accounting 
for the differences of the dispersion relation between an $e^-$-$p$- and 
an $e^-$-$e^+$-plasma), we find the acceleration rate due to Alfv\`en
waves,

\begin{equation}
\left( {d\gamma \over dt} \right)_{A} = 
{\pi \, (q - 1) \over q } \Omega_0^{2 - q} k_{min}^{q - 1} \> 
\left( {\delta B \over B_0} \right)^2 \, v_a^2 \, c^{q - 3} \> 
(\gamma \beta)^{q - 1},
\end{equation}
and the energy dispersion rate

\begin{equation}
\left( {d [\Delta\gamma]^2 \over dt} \right)_{A} = 
2 \, {\pi \, (q - 1) \over q \, (q + 2)} \Omega_0^{2 - q} 
k_{min}^{q - 1} \> \left( {\delta B \over B_0} \right)^2 \, v_a^2 
\, c^{q - 3} \> \gamma^q \, \beta^{q + 1}.
\end{equation}

\subsection{Energy dispersion due to inverse-Compton scattering}

Being a stochastic process, also inverse-Compton scattering
leads to energy dispersion. The respective energy-dispersion 
rates for Compton scattering in the Thomson regime are 
calculated in Appendix A.3.

For relativistic particles, the energy-dispersion rate due to
inverse-Compton scattering is $d(\Delta\gamma)^2 / dt \propto
\gamma^4$ and linearly dependent on 
$\int_0^{\infty} d\epsilon \, \epsilon^2 \, n_{ph} (\epsilon)$, 
i. e. the photon densities and the average of the square of
the photon energies. Their extremely strong dependence on 
particle energy ($\propto \gamma^2$ and $\propto \gamma^4$, 
respectively) implies that the energy-exchange and dispersion 
rates due to inverse-Compton scattering will usually dominate 
the total rates for very high particle energies. 

Which process is dominant at low particle energies,
depends critically on the particle density and the strength
of magnetohydrodynamic turbulences. In the case of high
densities, elastic scattering will govern the evolution of
these particles, while in the case of lower density and strong
plasma wave turbulences, wave-particle interactions (reacceleration)
and adiabatic losses will dominate their behavior.

Figs. 1 and 2 show the energy-exchange and energy-dispersion
rates for the different processes for a plasma of density
$3 \cdot 10^6 \, {\rm cm}^{-3}$, having a narrow distribution
around $\langle\gamma\rangle \approx 30$. A magnetic field
of $B = 0.1$~G is assumed and the blob is located at 
$z = 10^{-3}$~pc above an accretion disk with total luminosity
$L = 10^{45} \, {\rm erg \, s}^{-1}$ and moves outward with
bulk Lorentz factor $\Gamma = 15$. The magnetic-field
is assumed to be isotropized ($b = 2$). These parameter 
values are similar to those which result from the cooling 
of an ultrarelativistic pair plasma producing an X-ray and
$\gamma$-ray spectrum consistent with the observed spectrum
of PKS~0208-512.

\subsection{Pair annihilation losses}

A detailed discussion of the effects of pair annihilation on a
relativistic pair plasma in AGN jets can be found in B\"ottcher 
\& Schlickeiser (1996). In order to calculate the total pair
annihilation rate, we use the expression given by Svensson
(1982) which, for completeness, is quoted in Appendix A.4.
For relativistic particles, it declines roughly as $\dot n_{\pm}
(\gamma_{\pm}) \propto (\gamma_{\pm} \langle\gamma_{\mp}\rangle)^{-1}$,
where $\langle\gamma_{\mp}\rangle$ is the average Lorentz
factor of electrons or positrons, respectively, and becomes only 
important in the case of very high particle densities.

\section{Simulations of the jet evolution}

Since several energy loss and dispersion rates depend on the 
present particle distributions, Eq. (1) in its general form
represents a set of coupled integro-differential equations 
which, in general, can not be be solved analytically. For the 
purpose of numerical integration, we used an explicit code of 
finite differencing. Because the numerical simulations are extremely
time-consuming we tried to find simplified approximative
expressions for the coefficients in Eq. (1) in a way that the 
problem could be solved analytically. This is possible if 
both the cooling rate and the energy-dispersion rate are
dominated by inverse-Compton scattering for all occupied energy
states, which is the case if the pair distributions still have
a relativistic low-energy cutoff at $\gamma \gtrsim 10$. Then pair 
annihilation is negligible and provided that the cooling time scale 
is much shorter than the time scale of dilution of the pair plasma 
due to expansion, the source term in Eq. (1) may be neglected. 
Furthermore, the energy-loss and dispersion coefficients are dominated 
by synchrotron emission and inverse-Compton scattering and can very 
well be approximated by

\begin{equation}
{d\gamma \over dt} \approx - A_1 \, \chi(t) \, \gamma^2,
\end{equation}
\begin{equation}
{d (\Delta\gamma)^2 \over dt} \approx A_2 \, \chi(t) \, \gamma^4
\end{equation}

where $A_i$ are constants and both Fokker-Planck coefficients
have the same explicit time dependence $\chi(t)$.
For the resulting differential equation we found an approximative
analytical solution which is derived in Appendix B. Given a 
known distribution $n(\gamma_0, 0)$ at $t = 0$, we can forward
it in time to $t = t_1$ by

\begin{equation}
n(\gamma, T_1) = {e^{{A_1 \over A_2 \gamma} - {A_1^2 \over 2 A_2} 
\, T_1} \over 2 \, \gamma^3 \, \sqrt{ \pi \, A_2 \, T_1 / 2}}
\int\limits_1^{\infty} d\gamma_0 \> n(\gamma_0, 0) \, \gamma_0 \, 
e^{-{A_1 \over A_2 \gamma_0}} e^{- \left( {1 \over \gamma_0} - 
{1 \over \gamma} \right)^2 / (2 \, A_2 \, T_1) }
\end{equation}
where 

\begin{equation}
T := \int\limits_0^t dt' \> \chi(t').
\end{equation}
Since the explicit time dependence $\chi(t)$ of the coefficients,
being governed by the evolution of the dominant soft photon
field, is difficult to separate in realistic situations, we
choose time steps $t_1$ in a way that we may assume the 
coefficients $A_1 \chi(t)$ and $A_2 \chi(t)$ to be constant 
within this time interval, i. e. we approximate $\chi(t)$ by 
a step function. We use the solution (16) as long as the exact 
energy loss and dispersion rates of the lowest-energetic particles
do not deviate more than 10~\% from the representations (14) and 
(15). Comparison with numerical integrations of Eq. (1) showed
good agreement. After this phase we continue with numerical 
integrations of the full Fokker-Planck equation.

It is obvious that we cannot extend the simulations using the 
code described above to subrelativistic pair temperatures 
since especially the bremsstrahlung cross section which we
use is only valid in the ultrarelativistic limit.
However, under a wide range of assumptions, the pair 
plasma attains a quasi-thermal distribution whose temperature
depends basically on the energy density of the external
radiation field, the pair density, the magnetic field and 
the amplitude of plasma wave turbulences. 

Figs. 3 --- 6 show two examples of this evolution as
results of the numerical scheme described above. These
examples illustrate the general tendency towards the 
establishment of a quasi-thermal distribution inside
the jet component. At high particle energies the
real quasi-equilibrium distributions are truncated,
if radiative energy-loss processes dominate over
adiabatic losses (Fig. 4). If adiabatic losses
dominate, the high-energy end of the particle spectrum 
tends to be slightly harder than a thermal distribution
(Fig. 6). In both examples, the initial conditions 
were chosen in a way that the $\gamma$-ray compactness 
is low, which results in a time-averaged power-law 
$\gamma$-ray spectrum of photon number index 
$\alpha_{\gamma} = (s + 2)/2$ due to inverse-Compton 
scattering of soft photons, where $s$ is the spectral
index of the electron energy distribution.

In the case of extremely high particle densities ($n \ugr 
10^8$~cm$^{-3}$), quasi-thermalization is a consequence of 
elastic scattering becoming the dominant energy exchange 
process. One example for this fact is illustrated in 
Figs. 3 and 4. Fig. 4 demonstrates that for such high 
densities, the truncation of the distribution functions 
with respect to a thermal distribution at high energies 
is of minor importance.

For the case of lower densities, Schlickeiser (1985) 
de\-monstrated that if the spectral index $q$ of the
Alfv\`en wave turbulence equals 2, the combined action 
of stochastic acceleration and radiative losses also 
establishes a quasi-thermal distribution. Its temperature 
is given by 

\begin{equation}
\Theta = \sqrt{1 + (3 + a)^2 {\tilde p}_c^2} - 1
\end{equation}
where ${\tilde p}_c \, m_e c$ is the particle momentum 
for which the net energy loss or gain, respectively, 
vanishes, and $a = \alpha_1 / \alpha_2$ 
is the ratio of the coefficients governing the 
stochastic acceleration and energy-dispersion rates via 
$dp / dt = \alpha_1 p$ and $d(\Delta p)^2 / dt = \alpha_2
p^2$. The equilibrium temperature in Eq. (18) obviously
depends very sensitively on the efficiency of radiative 
cooling and can become quite high if the external photon
energy density is low, because the energy dependence of
adiabatic ($\propto \gamma$) and bremsstrahlung ($\propto 
\gamma^{1.1}$) energy losses is much weaker than that of 
inverse-Compton losses, leading to a very high equilibrium 
Lorentz factor (obviously, elastic-scattering [$\approx$~const.] 
cannot balance acceleration at all).

In our simulations, we find a similar behavior also for
the case of Kolmogorov turbulence, i. e. for $q = 5/3$.
Figs. 5 and 6 illustrate one example of this evolution. 
In this case, however, the deviation from a maxwellian 
distribution at high energies is more pronounced than 
in the case of a dense plasma. Assuming that inverse-Compton
scattering of accretion disk radiation is the dominant
cooling mechanism, which is balanced by stochastic
reacceleration, and that quasi-thermalization occurs at
relativistic temperatures, $\Theta \gg 1$, we find for the 
quasi-equilibrium temperature

\begin{equation}
\Theta_{eq} \approx 3 \cdot 10^5 \, B_0^7 \, R_{15}^{-2} \, 
n_e^{-3} \, \delta_{-1}^6 \, {z_{pc}^6 \over L_{46}^3},
\end{equation}
where $B_0$ is the magnetic field in Gauss, $R_{15}$ is the
blob radius in units of $10^{15}$~cm, $n_e$ is in cm$^{-3}$,
$(\delta B / B_0) = 0.1 \, \delta_{-1}$, $z_{pc} =
z / (1 \, {\rm pc})$, and $L_{46} = L_D / (10^{46} \, {\rm erg
\, s}^{-1})$. In order to derive this estimate, we have used 
Eqs. (32), (37) and (38) in the limit $\Theta \gg 1$.

The X-ray and $\gamma$-ray emission of a 
quasi-thermal, mildly relativistic plasma is dominated by 
its pair annihilation and pair bremsstrahlung radiation. 
For pair temperatures $\Theta \ukl 3$ the radiative 
output near the high-energy cut-off of the photon spectra 
is dominated by pair annihilation radiation which is 
insensitive to the pair distribution near its high-energy 
tail. The X-ray spectrum, to which only pair bremsstrahlung 
radiation contributes significantly and which has a photon
number spectral index of $\alpha \approx 1.1$, is equally 
insensitive to the detailed shape of the pair distribution 
at high energies. For this reason, the observeable radiation
from a mildly relativistic quasi-thermal pair plasma
may well be approximated by the respective emission 
of a thermal plasma whose distribution coincides with
the exact one up to the mean particle energy.

This situation changes if the plasma maintains a highly
relativistic temperature ($\Theta \gtrsim 5$). In this
case, the luminosities in synchrotron, SSC, and/or 
external Compton radiation may well dominate over 
pair bremsstrahlung and pair annihilation radiation, 
and the exact pair distribution has to be used in 
order to calculate the emitted X-ray and $\gamma$-ray
spectrum.

The pair annihilation rate in a thermal plasma can be 
aproximated by

\begin{equation}
\dt n = - { 2 \, \pi \, c \, r_e^2 \, n^2 \over (1 + \Theta) }
\> \left( {1 \over 1 + 6 \Theta } + { \Theta \over 0.25 +
\ln ( 1 + 1.1545 \Theta ) } \right)^{-1}
\end{equation}
(Svensson 1982). The bremsstrahlung emission of a thermal 
pair plasma can be calculated as

\begin{equation}
\dot n_{br} (k) = \sqrt{ 2 \over 3 \, \pi} { 2 \, c \, \alpha \, \st
\,  n^2 \over k \, \sqrt{\Theta} } e^{ - k / \Theta} g (k, \Theta)
\end{equation}
where $g (k, \Theta)$ are the respective gaunt factors for $e$-$e$ and
$e^+$-$e^-$ bremsstrahlung for which we use the approximations for 
transrelativistic pair plasmas found by Skibo et al. (1995).
The pair annihilation spectrum emitted by a thermal plasma is

\begin{equation}
\dot n_{ann} (k) = {c \, n^2 \over \Theta K_2^2 \left({1 \over \Theta}
\right)} e^{-(2 \, k^2 + 1)/(2 \, k \, \Theta)} \>
\int\limits_1^{\infty} d\gr \, (\gr - 1) \, e^{- \gr / (2 \, k \,
\Theta) } \sigma (\gr)
\end{equation}
(Dermer 1984) where

\begin{equation}
\sigma(\gr) = {\pi \, r_e^2 \over 1 + \gr} \>
\left[\left( {\gr^2 + 4 \gr + 1 \over \gr^2 -1 } \right)
\, \ln \left( \gr + \sqrt{\gr^2 -1} \right) - {\gr + 3 \over
\sqrt{\gr^2 -1}}\right].
\end{equation}

The bulk Compton process is calculated in the Thomson
regime, including the full angle-dependence of the external
radiation field (see BMS for details). 

We will now derive an estimate for the density required for
elastic scattering to dominate the effect of stochastic
reacceleration. To this aim, we compare the energy-loss due
to elastic scattering for particles of energy $\gamma \gg 
\langle\gamma\rangle$ to the respective reacceleration rate.
Using Eq. (2), we find

\begin{equation}
\left( {d\gamma \over dt} \right)_{Mo, \gamma \gg \langle\gamma\rangle}
\approx 6 \cdot 10^{-9} {n_4 \over \langle\gamma\rangle} \>
{\rm s}^{-1}
\end{equation}
where $n_4 = n_e / (10^4 \, {\rm cm}^{-3})$. The stochastic 
acceleration rate according to Eq. (12) is

\begin{equation}
\left( {d\gamma \over dt} \right)_{A} \approx 5 \cdot 10^{-2}
\, B_0^{7/3} \, \delta_{-1}^{2} \, n_4^{-1} \, R_{15}^{-2/3}
\, \langle\gamma\rangle^{-1} \> \gamma^{2/3} \> {\rm s}^{-1}.
\end{equation}
This yields the condition

\begin{equation}
n_e \> \ugr \> 3 \cdot 10^7 \, B_0^{7/6} \, \delta_{-1} \, 
R_{15}^{-1/3} \> \gamma^{1/3} \> {\rm cm}^{-3}
\end{equation}
for elastic scattering to dominate the energy loss at particle
energy $\gamma \gg \langle\gamma\rangle$. If this condition is 
fulfilled for particles at the high-energy tail of the pair 
distribution, reacceleration is inefficient as compared to elastic 
scattering. In this case, radiative and adiabatic cooling will 
prevent the particle distributions from attaining a stationary 
state.

\section{Evolution of a thermal plasma}

As we have shown in Section 3, in the case of very high 
particle densities ($n \ugr 10^8$ cm$^{-3}$), the pair 
plasma of a relativistic jet component approaches
a quasi-thermal distribution due to the dominant action of
elastic scattering. The evolution of a thermal distribution
of particles, subject to pair annihilation, bremsstrahlung, 
inverse-Compton, synchrotron, SSC, and adiabatic losses, 
can be followed using the expressions for the respective 
cooling rates of thermal pair plasmas. 

For this purpose, we assume that the thermalization timescale
(due to elastic scattering) is much shorter than the energy
loss timescale. In this case, M\o ller and Bhabha scattering
will always establish a thermal distribution,

\begin{equation}
 dn (\gamma) = {n \, \beta \gamma^2 \over \Theta \, K_2 \left( {1 \over
\Theta} \right) } e^{-{\gamma \over \Theta}} \, d\gamma,
\end{equation}
where $\Theta = { k_B T \over m_e c^2}$ is the dimensionless
temperature and $K_2$ is the modified Bessel function of second
kind of order 2. In order to determine the respective cooling
rates in a thermal plasma, we note first that

\begin{equation}
\left\langle p^2 \right\rangle = (m_e c)^2 \left\langle \gamma^2 - 1
\right\rangle =  (m_e c)^2 \cdot 3 \, \Theta {K_3 \left( {1 \over
\Theta} \right) \over K_2 \left( {1 \over \Theta} \right) }
\end{equation}
where $\langle . \rangle$ denotes the average over the distribution
(27). Let $W = {1 \over n} {dE \over dV}$ be the average particle
energy. Then, the cooling rate due to synchrotron losses is

\begin{equation}
\left( \dt W \right)_{\rm SY} \approx - c \, \st {B^2 \over 2 \, \pi} \, 
\Theta {K_3 \left( {1 \over \Theta} \right) \over K_2 \left( {1 \over 
\Theta} \right) } e^{-{\gamma_R \over \langle\gamma\rangle}}
\end{equation}
where the exponential is an approximate expression to include the
suppression of cooling due to the Razin-Tsytovich effect. 
$\langle\gamma\rangle = K_3(1/\Theta) / K_2 (1/\Theta) - \Theta$ 
is the mean Lorentz factor of the pairs (see eq. [38]) and 

\begin{equation}
\gamma_R \approx 2.1 \cdot 10^{-3} \, \left( {n \over {\rm cm}^{-3}}
\right)^{1 \over 2} \left( {B \over {\rm G}} \right)^{-1} \langle
\gamma\rangle^{-{1 \over 2}}
\end{equation}
is the Razin-Tsytovich Lorentz factor (Crusius \& Schlickeiser 1988).
The cooling rate due to inverse-Compton scattering of accretion
disk photons in the Thomson limit is

\begin{equation}
\left( \dt W \right)_{\rm IC} = - m_e c^2 \, {4 \over 15} {\pi^5 \,
\sigma_T \over c^2} \, \Gamma^2 \, 
\left( {m_e c^2 \over h} \right)^3\Theta \, {K_3 \left( 
{1 \over \Theta} \right) \over K_2 \left( {1 \over \Theta} \right)}
\int\limits_{R_{in}}^{R_{out}} dR \> R\, \Theta_D^4 (R) \, 
{ (x - \beta_{\Gamma} z)^2 \over x^4}
\end{equation}
where $R$ is the distance of a point in the accretion disk to the
centre, $R_{in}$ and $R_{out}$ are the boundary radii of the disk, 
$\Theta_D (R)$ is the dimensionless temperature of the accretion 
disk material at radius $R$, and $x = \sqrt{R^2 + z^2}$. If the 
point source approximation ($z \gg R_{out}$) holds, Eq. (31) 
reduces to

\begin{equation}
\left( \dt W \right)_{\rm IC} = - {\st \> L_0 \> \Theta \over \pi \,
\Gamma^2 (1 + \beta_{\Gamma})^2 z(t)^2} {K_3 \left( {1 \over
\Theta} \right) \over K_2 \left( {1 \over \Theta} \right) }.
\end{equation}
Fig. 7 illustrates that for a typical Shakura-Sunyaev disk of
total luminosity $L = 10^{46}$~erg~s$^{-1}$, surrounding a black
hole of mass $M = 10^8 \, M_{\odot}$, the point source approximation
is an appropriate choice for $z \ge 0.1$~pc. The temperature ($\Theta
= 2$ in Fig. 7) enters both Eqs. (31) and (32) in the same way and
does therefore not have any influence on the validity of the point
source approximation.

The cooling rate due to pair bremsstrahlung emission of a thermal pair
plasma has been calculated by Haug (1985b). We interpolate between
his expressions for the extreme-relativistic and the non-relativistic
regime:

\begin{equation}
- \left( \dt W \right)_{\rm br} =
\cases{ {{128}\over {3\, \sqrt{\pi}}} \, \alpha r_e^2 \, m_e c^3 \, n
\sqrt{\Theta} \quad & for $\ \Theta \ll 1$,\cr\cr
96 \, \alpha \, r_e^2 \, m_e c^3 \, n \, \Theta \, (\ln (2\Theta) +
0.673) \quad & for $\ \Theta \ugr 1$.\cr}
\end{equation}

For synchrotron-self-Compton scattering, we approximate the
synchrotron radiation field to be distributed over a small range
of energies and use its energy density

\begin{equation}
u_{sy} = {3 \, R_B \, n \over 4 \, c} \left\vert \dt W \right\vert_{SY}
\end{equation}
leading to

\begin{equation}
\left( \dt W \right)_{\rm SSC} = - {4 \over 3} c \st \, u_{sy} \>
{\left\langle \gamma^2 - 1 \right\rangle \over n}
= - {3 \over 2} {c \, \st^2 \, R_B \, n \, B^2 \over \pi} \> \Theta^2
\left( {K_3 \left[ {1 \over \Theta} \right] \over K_2 \left[ {1 \over
\Theta} \right]} \right)^2 \, e^{-{\gamma_R \over \langle\gamma\rangle}}.
\end{equation}
Using Eq. (8) for the adiabatic cooling rate, we find

\begin{equation}
\left( \dt W \right)_{\rm ad} = - {3 \over 2} \, m_e c^2 \,
b \, c \, {\bg \, \Gamma \over z} \, \Theta.
\end{equation}
The heating rate due to stochastic interactions with Alfv\`en waves, 
according to Eq. (12), is

\begin{equation}
\left( \dt W \right)_A = m_e c^2 \> \sqrt{\pi} \> {q - 1 \over q + 2} 
\, {\Gamma\left( {q + 4 \over 2} \right) \over q} \, 
2^{\nu} \, \Theta^{\nu - 2} \, {K_{\nu} \left( {1 \over \Theta}
\right) \over K_2 \left( {1 \over \Theta} \right)} \, \Omega_0^{2 - q} \,
k_{min}^{q - 1} \, \left( {\delta B \over B_0} \right)^2 \> v_a^2 \, 
c^{q - 3}
\end{equation}
where $\nu = (q + 3)/2$.

The mean kinetic energy of the electrons and positrons in a thermal
plasma of temperature $\Theta$ is

\begin{equation}
W_{\rm th} = {m_e c^2 \over \Theta \, K_2 \left( {1 \over \Theta}
\right) } \int\limits_0^{\infty} d\gamma \, (\gamma - 1) \, \gamma
\sqrt{\gamma^2 - 1} \; e^{-{\gamma \over \Theta}}
= m_e c^2 \, \left( { K_3 \left[ {1 \over \Theta} \right] \over K_2
\left[ {1 \over \Theta} \right] } - \Theta - 1 \right)
\end{equation}
(which in the non-relativistic limit approaches $W_{th} = m_e c^2 \left[
{3 \over 2} \Theta + {15 \over 2} \Theta^2 \right]$). Thus,

\begin{equation}
{\partial \Theta \over \partial W} = {1 \over m_e c^2} \> \left(
-1 + {1 \over \Theta^2} + {5 \over \Theta} {K_3 \left[ {1 \over \Theta}
\right] \over K_2 \left[ {1 \over \Theta} \right] } - \left[ K_3
\left( {1 \over \Theta} \right) \over \Theta \, K_2 \left( {1 \over
\Theta} \right) \right]^2 \right)^{-1},
\end{equation} 
and the respective cooling rates are determined by

\begin{equation}
\dt {\Theta} = {\partial \Theta \over \partial W} \cdot \dt W.
\end{equation}

The annihilation rate in a thermal plasma is well approximated by
Eq. (20). We found that numerical solutions of the Fokker-Planck 
equation in the case of a thermal pair plasma are in good agreement 
with the calculation described in this section as long as the pair 
temperature is $\Theta \ugr 4$. At lower temperatures cooling of the 
thermal plasma is slower than predicted by our Fokker-Planck treatment. 
We attribute this mainly to the ultrarelativistic approximation 
used for the pair bremsstrahlung process which is not an appropriate 
approach for mildly relativistic temperatures.

In Fig. 8, we show three examples of the evolution of a thermal
pair plasma in an AGN jet with different initial densities. Plasmas
of very high density end up with a density evolution which is almost
independent of density, since initially the density evolution is
dominated by pair annihilation losses which are proportional to
the square of the density. At lower densities, the dilution
of the plasma due to the expansion of the jet always dominates
the decline of the density, which thus always remains well below
the density of a plasma injected with high initial density. 
The temperature evolution depends only weakly on the density
because it is governed by adiabatic losses. The differences in
the temperature evolution curves are a consequence of the density
dependence of bremsstrahlung losses and of the heating rate via 
$v_a^2 \propto n^{-1}$.

The x-ray and $\gamma$-ray emission of a dense, mildly relativistic
pair plasma in the outer parts of an AGN jet is calculated in the
same way as for the dilute thermalized pair plasma, as described 
at the end of the previous section.

\section{Application to PKS 0208-512}

The quasar PKS~0208-512 is the most prominent example of 
MeV blazars, i. e. blazars exhibiting variable emission around 
several MeV and reaching extremely high fluxes in this frequency
range, often dominating the total energy output of the source.
Analyzing COMPTEL data of Phase I and II, Blom et al. (1995) 
found that during Phase I, PKS~0208-512 did not show a significant
signal in the COMPTEL energy range, while in Phase II, a pronounced
bump at 1 -- 3 MeV was detected. Quasi-simultaneous EGRET observations
revealed a strong signal from PKS~0208-512 during both Phase I and
II. However, in Phase II, EGRET detected the source with lower
significance than during Phase I. The EGRET spectra appear to be
harder during Phase I than in Phase II. During Phase III PKS~0208-512
is again detected by EGRET at a high flux level, whereas the
COMPTEL data for that period do not show evidence for emission from
PKS~0208-512 (Blom 1996). The variability in the EGRET range correlates
with spectral index, which itself varies more strongly than in the 
case of the average blazar (see Stacy et al. 1996 and Pohl et al. 1997). 
In fact, the flux level at 100 MeV stays approximately constant while 
the emission above 300 MeV is highly variable.

The variability timescale of the high-energy component of the
$\gamma$-ray spectrum of our model system is determined by the
light travel time through the plasma component and corresponds 
to $\sim 6$~h in the observer's frame. The timescales of 
quasi-thermalization and of dilution of the pair plasma 
due to pair annihilation and the expansion of the jet are
of the order of a few weeks. An outburst at MeV energies 
due to transrelativistic, quasi-thermalizing pair 
plasma in a relativistic jet is therefore generally not 
expected to be correlated with a $\gamma$-ray outburst observed 
by EGRET within the same observing period. In general terms, 
the flux in the EGRET range reflects the injection rate of 
high-energy pairs, whereas the flux of X-ray and soft
$\gamma$-ray photons is a measure of the injection rate 
of pairs at all energies averaged over several weeks. 

The instantaneous pair annihilation and pair bremsstrahlung 
radiation of one single thermalizing pair plasma blob is too 
weak to produce the observed MeV bump in the spectrum of 
PKS~0208-512. Therefore, if the MeV blazar phenomenon
is to be explained by pair annihilation radiation of cooled,
thermalized pair plasma in a relativistic jet, this jet must
be quasi-continuously filled with pair plasma. 

If, in contrast, the MeV bump is produced by inverse Compton
scattering of external radiation by the quasi-thermal, transrelativistic
pair plasma in the jet, reacceleration of particles must be efficient 
because else the X-ray and soft $\gamma$-ray spectrum would just be the
extension of the $\gamma$-ray power-law, resulting from Comptonization
of soft radiation in the Compton-cooled nonthermal plasma. The
bulk Compton process is therefore mainly relevant in the low to 
medium density case, in which relativistic temperatures can be 
maintained. However, a problem with this interpretation is that
the quasi-equilibrium temperature of the pair plasma depends
extremely sensitively on the magnetic field strength, the level
of hydromagnetic turbulence and the injection height of the
plasma blob, as indicated by Eq. (19). Therefore, if inverse-Compton
scattering of external radiation by quasi-thermal pair plasma
in the jet plays an important role, we would expect to see
bulk-Compton bumps in different objects at a largely different 
frequencies, for which no evidence has yet been found. There is
no obvious reason why this bump should be produced preferrably
at several MeV.

The dashed line in Fig. 9 shows a fit to the EGRET spectrum during
Phase II (VP 220) where we adopted the following set of parameters:
$n_e = 5 \cdot 10^5 \, {\rm cm}^{-3}$, $R_B = 6 \cdot 10^{15}$~cm, 
$\gamma_{1\pm} = 500$, $\gamma_{2\pm} = 2 \cdot 10^4$, $s = 2.4$,
$B_0 = 0.7$~G, $b = 2$, $z_i = 2 \cdot 10^{-3}$~pc, $\Gamma = 10$,
$\theta_{obs} = 6^{\circ}$, $L_0 = 10^{46} \, {\rm erg \, s}^{-1}$,
$M_0 = 10^{8} \, M_{\odot}$. The high-energy spectrum results
predominantly from inverse-Compton scattering of external
(accretion-disk) photons. The photon spectrum has been integrated 
over the cooling time of the ultrarelativistic pair population 
because this time is only several hundred seconds (in the observer's 
frame) which cannot be resolved by COMPTEL and EGRET. The fluence 
resulting from our time integration has been re-converted into an 
average flux by multipling by the repetition rate of subsequent 
blobs. Since during an observed time interval $\Delta t_{obs}$ 
the blob travels a distance $\Delta l = D \, \Gamma \, \beta_{\Gamma} 
\, c \, \Delta t_{\rm obs}$~cm, a jet filled quasi-continuously 
with relativistic material ($\Delta l \approx R_B$) corresponds 
to a repetition time in the observer's frame of $\Delta t_{\rm obs} 
\approx 2000$~s which translates to a quasi-continuous energy 
input into the jet of $L_{jet} \approx 5 \cdot 10^{43}$~erg~s$^{-1}$ 
during an EGRET outburst. 

The bump of the radiation from an accretion disk of 
$L = 10^{46}$~erg~s$^{-1}$ at redshift $z \approx 1$ 
corresponds to $\nu \, F_{\nu} \approx 10^{-6} \, 
{\rm MeV \, cm}^{-2} \, {\rm s}^{-1}$, if we assume $H_0 = 75 \, 
{\rm km / (s \cdot Mpc)}$, $q_0 = 0.5$. No simultaneous broadband 
measurements in the radio to UV frequency range are available. For 
this reason, we make no attempt to fit the synchrotron component. 
Anyway, the radio emission probably originates in a much larger 
volume than considered here, in accord with the fact that there 
is no correlation between the simultaneously observed \gamr flux 
and cm-radio flux of blazars (M\"ucke et al. 1996, 1997). 

Using the methods described in the previous sections, we follow 
the evolution of the plasma blob through the transrelativistic 
phase until it is diluted by pair annihilation and the expansion
of the jet to less than $10^{-4}$ of its initial density. Due to
the rather low initial density, quasi-thermalization occurs at
a temperature of $\Theta \approx 100$, implying that the X-ray
and soft $\gamma$-ray spectrum is strongly dominated by inverse
Compton scattering of accretion disk radiation.

In order to reproduce the MeV bump, we assume that some fraction
of the jet is filled with pair plasma, evolving in the same
way as the ultrarelativistic component, over a long period of
time. We find this fraction to be $\sim$~10~\%. This is consistent 
with the observation of strong variability in the EGRET regime 
if one assumes that during the quiescent phases the injection
of relativistic pair plasma into the jet is very inefficient,
and is in perfect agreement with the typical flaring duty cycle 
of EGRET detected blazars. The combined inverse-Compton, pair 
annihilation and pair bremsstrahlung spectrum emitted by the 
quasi-thermalized pair plasma jet as it moves out and is diluted 
due to jet expansion and pair annihilation, is shown by the solid 
line in Fig. 9. 

We point out that the MeV bump can be fit with a self-consistent
pair annihilation spectrum only under the assumption of a
cylindrical jet, which seems to contradict the assumption
that the gas pressure of the pair plasma greatly exceeds
the magnetic pressure, most probably responsible for the jet
confinement. 

\section{Summary and conclusions}

Motivated by the standard model for $\gamma$-ray emission from
blazars invoking jets filled with ultrarelativistic pair plasma
oriented at a small angle with respect to the line of sight, we 
investigated the further evolution of the leptonic material
inside such a jet after the initial phase in which the 
high-energy ($> 100$~MeV) $\gamma$-ray radiation is 
produced.

We demonstrated that either the action of elastic
scattering of particles off each other or the balance
of reacceleration by turbulent plasma waves to radiative
losses can, under a wide range of parameters, establish a
quasi-thermal particle distribution inside the jet,
irrespective of the initial pair distribution at the
time of injection. Extreme conditions (high density, 
very weak magnetic field, injection close to the 
accretion disk) are necessary to achieve mildly relativistic 
temperatures ($\Theta \sim 1$). Radiation from a quasi-thermal 
pair plasma can produce the temporary MeV bumps observed in 
MeV blazars. This can be accomplished most easily via
inverse-Compton scattering of external soft radiation
by a quasi-continuously filled jet. 

Alternatively, the MeV bump could be produced by pair
annihilation radiation, if cooling of the pair distribution
is very efficient and the dilution of the plasma through
jet expansion is negligible. We found that only the
unrealistic assumption of a cylindrical jet yields an 
acceptable fit to the observed MeV bump, if the pair
plasma producing this bump is subject to the same
processes (in the same environment) as the ultrarelativistic
pair plasma producing the EGRET spectrum.

In this picture, an MeV outburst reflects an increasing supply of
relativistic electrons and positrons into the jet on much longer
timescales than those typically observed in EGRET outbursts
at higher energies. It is only weakly or not at all correlated 
to activity in the EGRET range which corresponds to an increased
energy density at the acceleration site, leading to a harder
particle spectrum of the injected pairs.

If the bulk Compton process is the dominant mechanism 
for the production of the MeV bump in MeV blazars, then 
there is no obvious reason for a ``universality'' of the
bump photon energy at 3 --- 10~MeV. Mainly depending on the 
magnetic field strength, the level of hydromagnetic turbulence
and the injection height of the plasma blob above the accretion
disk, similar sources with bulk-Compton radiation bumps 
at different energies should exist as well.

In general, our treatment is also applicable to the so-called
hadronic jet models, where ultrarelativistic protons initiate
a pair cascade in the jet (Mannheim et al. 1991, Mannheim \& 
Biermann 1992, Mannheim 1993). However, although the jet simulation 
code of BMS is able to handle problems in which the blob is optically
thick to $\gamma\gamma$ pair production, it is clearly beyond 
the scope of this paper to follow the proton-initiated pair cascades 
in detail. Nevertheless, a few general conclusions may be drawn 
from the fact that in those models, the magnetic fields are fairly 
high and a considerable level of turbulence might be present. Our
simulations indicate that at magnetic fields of $B \gtrsim 10$~G,
as usually assumed in hadronic jet models, and the moderate
pair densities expected to result in the cascades, the pair
plasma will maintain a highly relativistic temperature, $\Theta 
\gtrsim 100$. Thus, no pair annihilation feature will be observable;
instead, strong synchrotron and SSC bumps at 
$\sim 3 \cdot 10^7 \cdot D/(1+z) \, (B/G) \, \Theta^2$~Hz and 
$\sim 4 \cdot 10^8 \cdot D/(1+z) \, (B/G) \, \Theta^4$~Hz,
respectively, would be expected from the thermalized cascade 
plasma. 

\acknowledgements{We thank the anonymous referee for valuable
comments and suggestions which led to significant improvements
of the manuscript. This work has been partially supported by 
NASA grant NAG~5-4055. R. S. acknowledges partial support by 
the DARA.}

\appendix

\section{Energy exchange and dispersion rates}
\subsection{Energy dispersion due to elastic scattering}

Following Dermer (1985), the energy dispersion rate 
is given by

$$
{d (\de)^2 \over d t} = 8 \pi^2 \, N_{\pm} \, c \, \int\limits_1^{\infty} 
d\gr \> (\gr^2 - 1) $$
\begin{equation} \;\;
\int\limits_1^{\infty} d\gc \> \sqrt{\gc^2 - 1} \int\limits_{-1}^1
d u \> f_1 (\game) \, f_2 (\gz) \> \sigma_{Mo} (\gr) \deqm
\end{equation}
where

$$ \deqm = {2 \, \pi \, r_e^2 \, \left(\gc^2 - 1\right) \over 
\sigma_{Mo} (\gr) \, \gcm^2 \, \left(\gcm^2 - 1 \right)} \> \cdot 
$$
$$ \cdot \Biggl( u^2 \> \left[{\gr^2 \over 2} - \left(2 \, \gcm^4 -
\gcm^2 - {1 \over 4} \right) (2\,\ln 2 - 1) + {(\gcm^2 - 1)^2 \over
12} \right] $$
$$ + {\left[1 - u^2 \right]\over 2} \, \Biggl[ \gr^2 \left({1 \over 2} 
\ln 2 + \ln\Lambda \right) 
$$
\begin{equation}
\hskip 1.5cm - \left(2 \, \gcm^4 - \gcm^2 - {1 \over 4}
\right) + {(\gcm^2 - 1)^2 \over 6} \Biggr] \Biggr).
\end{equation}
Here $\gr$ is the Lorentz factor of the relative motion of
the interacting particles, $\gc$ is the Lorentz factor (normalized
velocity: $\bc$) of motion of the center-of-momentum (cm) frame 
with respect to the blob frame, $\gcm$ is the Lorentz factor 
(normalized velocity: $\bcm$) of the particles in the 
center-of-momentum frame, and $u$ is the cm angle cosine
between $\bcmv$ and $\bcv$.

It is accurate within a few percent to retain only the 
leading terms $\sim u^2$ and $\sim (1 - u^2)$, respectively, 
that is the terms $\sim u^2 \, \gr^2 / 2$ and $\sim (1 - u^2) 
\, \gr^2 \, (\ln\sqrt 2 / 2 + \ln\Lambda) / 2$ in Eq. (6). 
Nayakshin and Melia (1998) suggested an approximation based on the
further neglect of all terms not containing an $\ln\Lambda$, arguing 
that this contribution is the one resulting from small-angle 
scatterings and that, at the point where other contributions 
may become important, the use of the Fokker-Planck equation 
is no longer a good approach since then large-angle scattering, 
changing the particles' energy considerably, dominate the 
energy exchange in elastic scattering events. Our simulations 
show that in the case of the system we are dealing with the 
energy distributions of the pairs remain relatively
narrow. In this case, the dispersion may indeed be approximated
by the expression found by Nayakshin \& Melia which ensures
the validity of the Fokker-Planck equation. Nevertheless,
we retain both contributions ($\sim u^2$ and $\sim [1 - u^2]$) 
in order to see whether large-angle scattering is important 
at the point where the plasma starts to thermalize 
(i. e. when elastic scattering becomes the dominant process).

With this simplification, the dispersion rate due to elastic 
scattering off an isotropic distribution of particles $n_2 
(\gamma_2)$ is given by

\begin{equation}
\left( {d [\Delta\gamma]^2 \over dt} \right)_{Mo} = 
\int\limits_1^{\infty} d\gamma_2 \> n (\gamma_2) \, D(\gamma, \gamma_2).
\end{equation}
where

\vbox{
$$ D (\gamma_1, \gz) = {4 \, \pi \, r_e^2 \, c \, n_2 \, \over \gt^2 \bt 
\, \gz^2 \bz} \Biggl\lbrace \left( \lnl + {[\gt - \gz]^2 \over 4}
\, [1 - \lnl] \right) \, \cdot $$
$$ \cdot \, \left( 2 \, \ln \left[ \gcm (1 + \bcm) \right] + 2 \, 
\gcm^2 \bcm - {1 \over \bcm} \right) $$
$$ + {(\gt + \gz)^2 \over 2} \left( \gcm^2 \bcm + {1 \over 2} \gcm -
\ln\left[ \gcm (1 + \bcm) \right] \right) $$
$$- \left( \gcm^4 + {3 \over 2} \gcm^2 \right) \bcm - {1 \over 
\bcm} + {5 \over 2} \ln\left( \gcm [1 + \bcm] \right) \Biggr\rbrace 
\Biggr\vert_{\gcm^{min}}^{\gcm^{max}}$$ 
\begin{equation}
\end{equation} }
\noindent
is the monoenergetic dispersion rate and

\begin{equation}
\gcm^{min/max} = \sqrt{ {1 \over 2} (1 + \gamma_1 \gz [1 \mp \be\bz] ) }.
\end{equation}

\subsection{Bremsstrahlung energy loss}

Using again the technique of Dermer (1985), we find the 
single-particle energy loss rate due to pair bremsstrahlung 
emission to be

$$ \left( {d \gt \over d t} \right)_{br} = - {\pi \, \Npm \, c \over
\bt \, \gt^2} \int\limits_1^{\gr^{max}} d\gr \> \gr^2 \, \br^2 $$
\begin{equation}
\;\;\;\;\;\;\;\;
\int\limits_{\gc^{min}}^{\gc^{max}} d\gc \> { f_2 \left( 2 \, \gcm \,
\gc - \gt \right) \over \gcm \bcm \, \gc + \gcm \, \gc - \gt} \> \kq
\end{equation}
where

$$ \kq \equiv \int\limits_0^{\ecm^{max}} d\ecm \> \ecm \left({d \sigma_{br} \over d k} \right)_{cm} = $$
$$ { \alpha \, \st \over \pi} \, \gcm \, \Biggl\lbrace \bcm \left(\ln\left[4 \gcm \over \bcm \right] - {1 \over 2} \right) \left( 4 - 2 \, \bcm + \bcm^2 \right) $$
$$ + (1 - \bcm)^2 \left(\ln[\gcm(1 - \bcm)] - {1 \over 2} \right) +
{\bcm^2 \over 3} \, (3 + \bcm) $$
$$ - (1 - \bcm) \left( 3 \ln[\gcm (1 - \bcm)] + 1 \right) +
3 \ln\gcm + {7 \over 6} $$
\begin{equation}
\;\;\;\;\;
+ (1 - \bcm)^3 \, \left(\ln[\gcm (1 - \bcm)] - {1 \over 3} \right)
\Biggr\rbrace \Biggr\vert_{\gcm = \gcm^{max}}
\end{equation}
(cf. DL). The maximum photon energy $\ecm^{max}$ is given by

\begin{equation}
\ecm^{max} = \gcm \, \bcm^2.
\end{equation}

\subsection{Energy dispersion due to inverse-Compton
scattering} 

The energy dispersion rate due to inverse-Compton 
scattering of photons of differential density 
$n_{ph} (\epsilon, \Omega_{ph})$ in the Thomson regime is

$$ \left( {d (\Delta\gamma)^2 \over dt} \right)_{EIC} = 4 \, \pi \, 
c \, \sigma_T \gamma^2 \, \beta^2 \left( \gamma^2 \, \beta^2 + 
{1 \over 6} \right) \cdot $$
\begin{equation}
\hskip 2cm \cdot \int\limits_{-1}^1 d\eta_{ph} 
\int\limits_0^{\infty} d\epsilon \> \epsilon^2 n_{ph} (\epsilon,
\Omega_{ph})
\end{equation}
where $\eta_{ph}$ is the cosine of the angle between soft photon
momentum and jet axis. Inserting the photon number density (I.8) 
yields

$$ \left( {d (\Delta\gamma)^2 \over dt} \right)_{EIC} = {48 \, \pi \,
\sigma_T \over c^2} \Gamma^3 \left( {m_e c^2 \over h} \right)^3
\zeta(5) \cdot $$
\begin{equation} \;\;\;\;\;\;\;\;
\cdot \, \gamma^2 \, \beta^2 \, \left( \gamma^2 \beta^2 + {1 \over 6}
\right) \int\limits_{R_i}^{R_a} dR \> \Theta^5 (R) {(x - \beta_{\Gamma}
z)^3 \over x^5}
\end{equation}
where $R_{i/a}$ is the radius of the inner/outer edge of the 
accregion disk, respectively, $x = \sqrt{R^2 + z^2}$ and 
$\zeta(n) = \sum_{k=1}^{\infty} k^{-n}$.

In the case of an isotropic soft photon field we have $n_{ph} (\epsilon, 
\Omega_{ph}) = n_{ph} (\epsilon) / 4\pi$ (e. g. synchrotron
photons in our model), and Eq. (19) reduces to

$$\left( {d (\Delta\gamma)^2 \over dt} \right)_{SSC} \!\!\!\!\!\!\!
= 2 \, c \, \sigma_T \, \gamma^2 \beta^2 \, \left( \gamma^2 
\beta^2 + {1 \over 6} \right) \int\limits_0^{\infty} d\epsilon \> 
\epsilon^2 \, n_{sy} (\epsilon).$$ 
\begin{equation}
\end{equation}

\subsection{Pair annihilation losses}

The general expression for the catastrophic 
particle losses due to pair annihilation can be written as

$$\left( {d \, n_{\pm} (\gamma, t) \over d t } \right)_{PA} \!\!\!\!\!\!
= - n_{\mp} (\gamma, t) \int\limits_1^{\infty} d\gmp \,
n_{\pm} (\gmp) \, \overline {v \sigma} (\gp, \gm)$$ 
\begin{equation}
\end{equation}
where

$$
\overline {v \sigma} (\gp, \gm) = {c \, \pi \, r_e^2 \over \beta_-
\gamma_-^2 \, \beta_+ \gamma_+^2 } \cdot
$$
\begin{equation}
\;\;\;\;\;\;\;\;
\cdot \left( \bcm^3 \gcm^2 L[\bcm] - 2 \, \gcm^2 + {3 \over 4} \,
L^2 [\bcm] \right) \biggr\vert_{\gcm^{min}}^{\gcm^{max}}
\end{equation}
is the angle-averaged reaction rate and

\begin{equation}
L(\beta) \equiv \ln \left( {1 + \beta \over 1 - \beta} \right)
\end{equation}
(Svensson 1982). $\gcm^{min}$, $\gcm^{max}$ are given by the
kinematic restrictions, namely

$$
\sqrt{{1 \over 2} \left( 1 + \game \gz [1 - \be \bz] \right)}
$$
\begin{equation}
\;\;\;\;\;\;\;\;\;\;\;\;\;\; \le \gcm \le \sqrt{{1 \over 2} \left( 1 +
\game \gz [1 + \be \bz] \right)}.
\end{equation}

\section{Approximative analytical solution}

As we found in section 4, for ultrarelativistic particles with
$\gamma \ugr 100$, and neglecting pair annihilation, the Focker-Planck
equation (1) can be approximated as

\begin{equation}
{1 \over \chi(t)} {\partial n \over \partial t} - {\partial \over 
\partial \gamma} \left( A_1 \, \gamma^2 \, n + {A_2 \over 2} {\partial 
\over \partial \gamma} \left[ \gamma^4 \, n \right] \right) = 0
\end{equation}

where $n = n(\gamma, t)$ and $\partial n / (\chi[t] \partial t) =
\partial n / \partial T$. Laplace transformation of Eq. (B1)
to

\begin{equation}
N(\gamma, s) = \int\limits_0^{\infty} dT \, n(\gamma, t) \, e^{s T}
\end{equation}
yields an ordinary differential equation of second order in $\gamma$:

\begin{equation}
\phi_0 (\gamma) \, N (\gamma) + \phi_1 (\gamma) \, N' (\gamma)
+ \phi_2 (\gamma) N'' (\gamma) = n(\gamma, 0)
\end{equation}
where

\begin{equation}
\phi_0 (\gamma) := s - 2 \, A_1 \, \gamma - 6 \, A_2 \, \gamma^2, 
\end{equation}
\begin{equation}
\phi_1 (\gamma) := - A_1 \, \gamma^2 - 4 \, A_2 \, \gamma^3, 
\end{equation}
\begin{equation}
\phi_2 (\gamma) := - {A_2 \over 2} \gamma^4.
\end{equation}

Eq. (B3) can be written in the form

\begin{equation}
\left( p[\gamma], N'(\gamma) \right)' - g(\gamma) \, N = - {n(\gamma,
0) \over f(\gamma)}
\end{equation}
with

\begin{equation}
p(\gamma) = e^{-{2 \, A_1 \over A_2 \, \gamma}} \, \gamma^8,
\end{equation}
\begin{equation}
f(\gamma) = {A_2 \over 2} e^{{2 \, A_1 \over A_2 \, \gamma}} \, \gamma^{-4},
\end{equation}
\begin{equation}
g(\gamma) = {2 \over A_2} \, e^{-{2 \, A_1 \over A_2 \, \gamma}}
\, \left( s \, \gamma^4 - 2 \, A_1 \, \gamma^5 - 6 \, A_2 \,
\gamma^6 \right)
\end{equation}

In the first step we search for a homogeneous solution $h(\gamma)$
which satisfies the equation

\begin{equation}
\left( p[\gamma] h' \right)' - g(\gamma) \, h = 0.
\end{equation}
Substituting

\begin{equation}
h(\gamma) = p(\gamma)^{-1/2} \, P(x),
\end{equation}
choosing $x = \gamma^{-1}$ and inserting the functions $g$ and $p$,
yields a differential equation for $P(x)$:

\begin{equation}
x \, P''(x) + 2 \, P'(x) - (\zeta \, x + \eta) \, P(x) = 0
\end{equation}
where

\begin{equation}
\zeta = {A_1^2 \over A_2^2} + {2 \, s \over A_2},
\end{equation}
\begin{equation}
\eta = {2 \, A_1 \over A_2}.
\end{equation}

Two linearly independent solutions of Eq. (B 13) are given by

\begin{equation}
\tilde P_1 (x) = {1 \over x} \> M_{- {\eta \over 2 \alpha}, \, 
{1 \over 2}} ( 2 \alpha \, x ),
\end{equation}
\begin{equation}
\tilde P_2 (x) = {1 \over x} \> W_{- {\eta \over 2 \alpha}, \,
{1 \over 2}} (2 \alpha \, x)
\end{equation}
where $\alpha = \sqrt{\zeta}$. Since we did not find an analytical
solution for the inverse Laplace transformation 

\begin{equation}
G(\gamma, \gamma_0, T) = {1 \over 2 \, \pi \, i} \int\limits_{x - i 
\infty}^{x + i \infty} d s \> e^{T s} G(\gamma, \gamma_0, s)
\end{equation}
of the Green's function resulting from the homogeneous solutions 
$h_{1/2} (\gamma) = p^{-{1/2}} (\gamma) \, \tilde P_{1/2} (1 /
\gamma)$ we simplify Eq. (B 13) in a convenient manner. Since the
energy loss and dispersion rates are dominated by the
inverse-Compton process, we find for the coefficients

\begin{equation}
A_1 \sim c \, \sigma_T \, \int\limits_0^{\infty} d\epsilon \> \epsilon
\, n_{ph} (\epsilon),
\end{equation}
\begin{equation}
A_2 \sim c \, \sigma_T \, \int\limits_0^{\infty} d\epsilon \> 
\epsilon^2 \, n_{ph} (\epsilon)
\end{equation}
where $\epsilon$ and $n_{ph} (\epsilon)$ are dimensionless energy and
spectral photon number of the accretion disk photon field in the
blob rest frame. Hence, we can estimate

\begin{equation}
{A_1 \over A_2} \sim {1 \over \langle \epsilon \rangle} \sim 10^5
\end{equation}
for the accretion disk model we used (see BMS, section 2.1).
Since for the Laplace transform $\Re s > 0$, we have $\vert \zeta 
\vert \ugr 10^{10}$ and $\eta \sim 10^5$. Thus, for particle energies 
$\gamma \ukl 10^4$, $\zeta x \gg \eta$, and we may neglect the term 
$\sim \eta$. The numerical values for the coefficients $A_1$ and $A_2$
which we find during our simulations confirm the above estimate.
Therefore, Eq. (B 13) is well approximated by

\begin{equation}
P''(x) + {2 \over x} P'(x) - \zeta \, P(x) = 0
\end{equation}
which is solved by

\begin{equation}
P_{1/2} (x) = {1 \over x} \, e^{\pm \alpha x}
\end{equation}
where, again $\alpha = \sqrt{\zeta}$. The homogeneous solutions 
are thus

\begin{equation}
h_{1/2} (\gamma) = e^{\left( {A_1 \over A_2 \gamma} \pm \alpha \right)
{1 \over \gamma}} \, \gamma^{-3}.
\end{equation}
The Laplace-transformed Green's function $G(\gamma, \gamma_0, s)$ is 
then constructed as

$$ G(\gamma, \gamma_0, s) = {1 \over C} \cdot $$
\begin{equation}
\cdot \, \Bigl\lbrace \Theta (\gamma_0 - \gamma) \, h_1 (\gamma)
\, h_2 (\gamma_0) + \Theta (\gamma - \gamma_0) \, h_2 (\gamma) \,
h_1 (\gamma_0) \Bigr\rbrace
\end{equation}
where

\begin{equation}
C = p (\gamma) \, W \lbrace h_1, h_2 \rbrace (\gamma) = 2 \, \alpha
\end{equation}
and $W \lbrace h_1, h_2 \rbrace$ denotes the Wronskian. The functions
$h_1$ and $h_2$ are chosen in a way that $h_1$ fullfills the boundary
condition at $\gamma \to 1$ and $h_2$ fullfills the boundary condition
for $\gamma \to \infty$ where we simply have to use the condition
of finite solutions $G(\gamma, \gamma_0, s)/f(\gamma_0)$ at $\gamma
\to \infty$ and $\gamma \ll \alpha$. (In the approach used here we
can also write Eq. (B1) differential in kinetic energy $E := 
\gamma - 1$ yielding the same solution in $E$ instead of $\gamma$; 
then the boundary condition is determined by a finite value of $G / f$
at $E = 0$.) Thus, the resulting Laplace transformed Green's 
function is

\begin{equation}
G(\gamma, \gamma_0, s) = - {e^{{A_1 \over A_2} \left( {1 \over \gamma} 
+ {1 \over \gamma_0} \right) } \over 2 \, \alpha \, \gamma^3 \, 
\gamma_0^3} \, e^{- \alpha \left\vert {1 \over \gamma_0} - {1 
\over \gamma} \right\vert}.
\end{equation}
The inverse Laplace transform of this Green's function is

$$ G(\gamma, \gamma_0, T) = $$
\begin{equation}
\;\;\;\;\; - {e^{ {A_1 \over A_2} \left( {1 \over \gamma}
+ {1 \over \gamma_0} \right)} \over 4 \, \gamma^3 \gamma_0^3} \,
{ A_2 \, e^{- {A_1^2 \over 2 \, A_2} \, T} \over \sqrt{ \pi \, A_2 \, 
{T \over 2} }} \, e^{-\left( {1 \over \gamma_0} - 
{1 \over \gamma} \right)^2 / (2 \, A_2 \, T)},
\end{equation}
and an approximative solution of Eq. (B 1) is given by

$$
n(\gamma, T) = - \int\limits_1^{\infty} d\gamma_0 \> G(\gamma,
\gamma_0, T) \, { n(\gamma_0, 0) \over f(\gamma_0)}
$$ 
$$
= {e^{{A_1 \over A_2 \gamma} - {A_1^2 \over 2 A_2} \, T} \over
2 \, \gamma^3 \, \sqrt{ \pi \, A_2 \, T / 2}} \, \cdot $$
\begin{equation}
\;\;\;\;\;\; \cdot \, \int\limits_1^{\infty}
d\gamma_0 \> n(\gamma_0, 0) \, \gamma_0 \, e^{-{A_1 \over A_2 \gamma_0}}
e^{- \left( {1 \over \gamma_0} - {1 \over \gamma} \right)^2 / (2 \, A_2
\, T) }.
\end{equation}

\eject

\begin{figure}
\rotate[r] {
\epsfxsize=12cm
\epsffile[100 20 600 50] {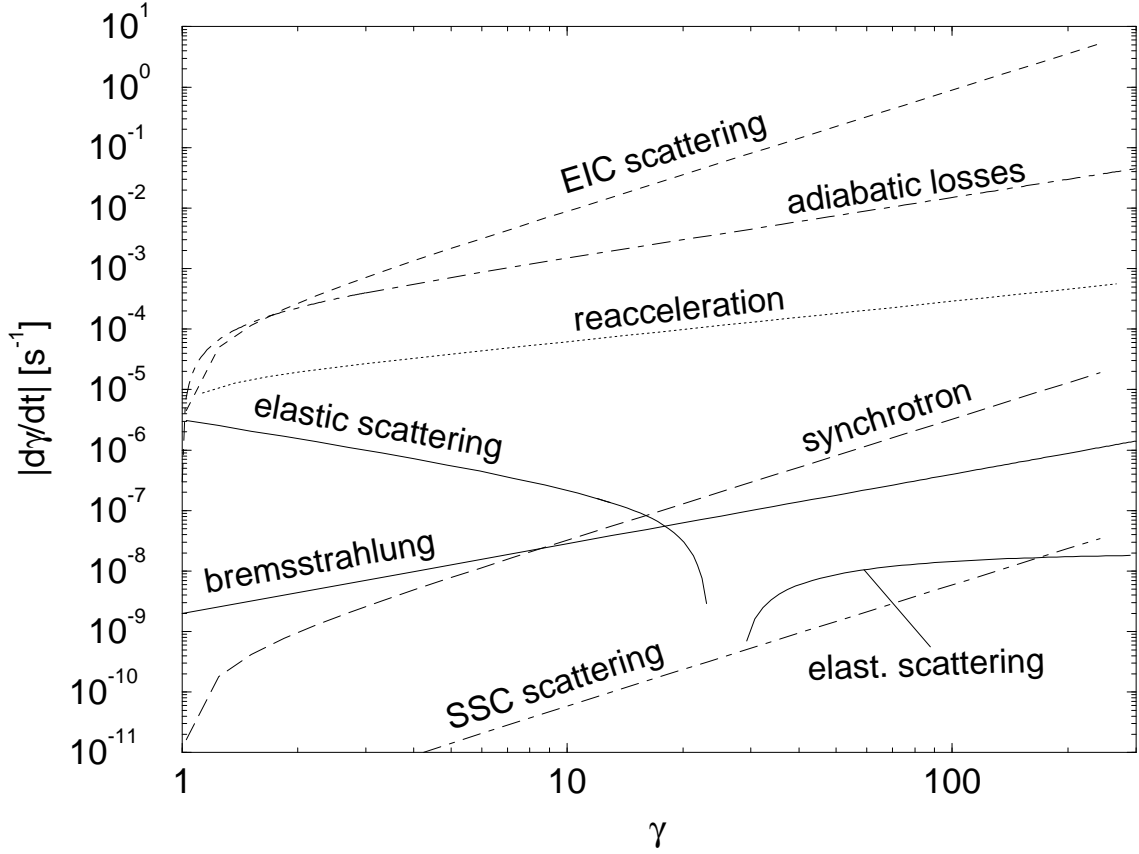} }
\caption[]{Energy-loss rates in a dense relativistic jet for
the following parameters: $n = 3 \cdot 10^6 \, {\rm cm}^{-3}$,
$\langle\gamma\rangle \approx 30$, $B_0 = 0.1$~G, $\delta B /
B_0 = 0.1$, $z = 10^{-3}$~pc, $L_0 = 10^{46} \, {\rm erg \, s}^{-1}$,
$\Gamma = 15$}
\end{figure}

\eject

\begin{figure}
\rotate[r] {
\epsfxsize=12cm
\epsffile[100 20 600 50] {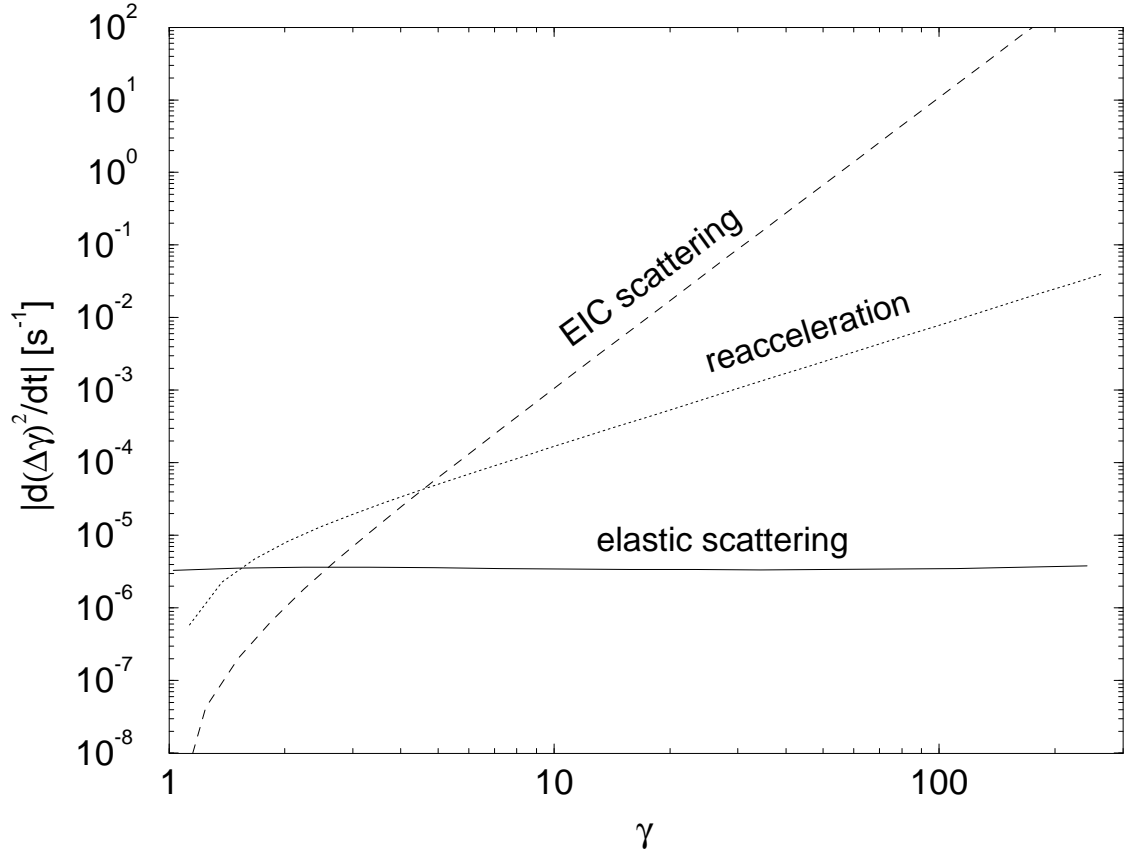} }
\caption[]{Energy-dispersion rates in a dense relativistic jet for
the same set of parameters as in Fig. 1}
\end{figure}

\eject

\begin{figure}
\rotate[r] {
\epsfxsize=12cm
\epsffile[100 20 600 50] {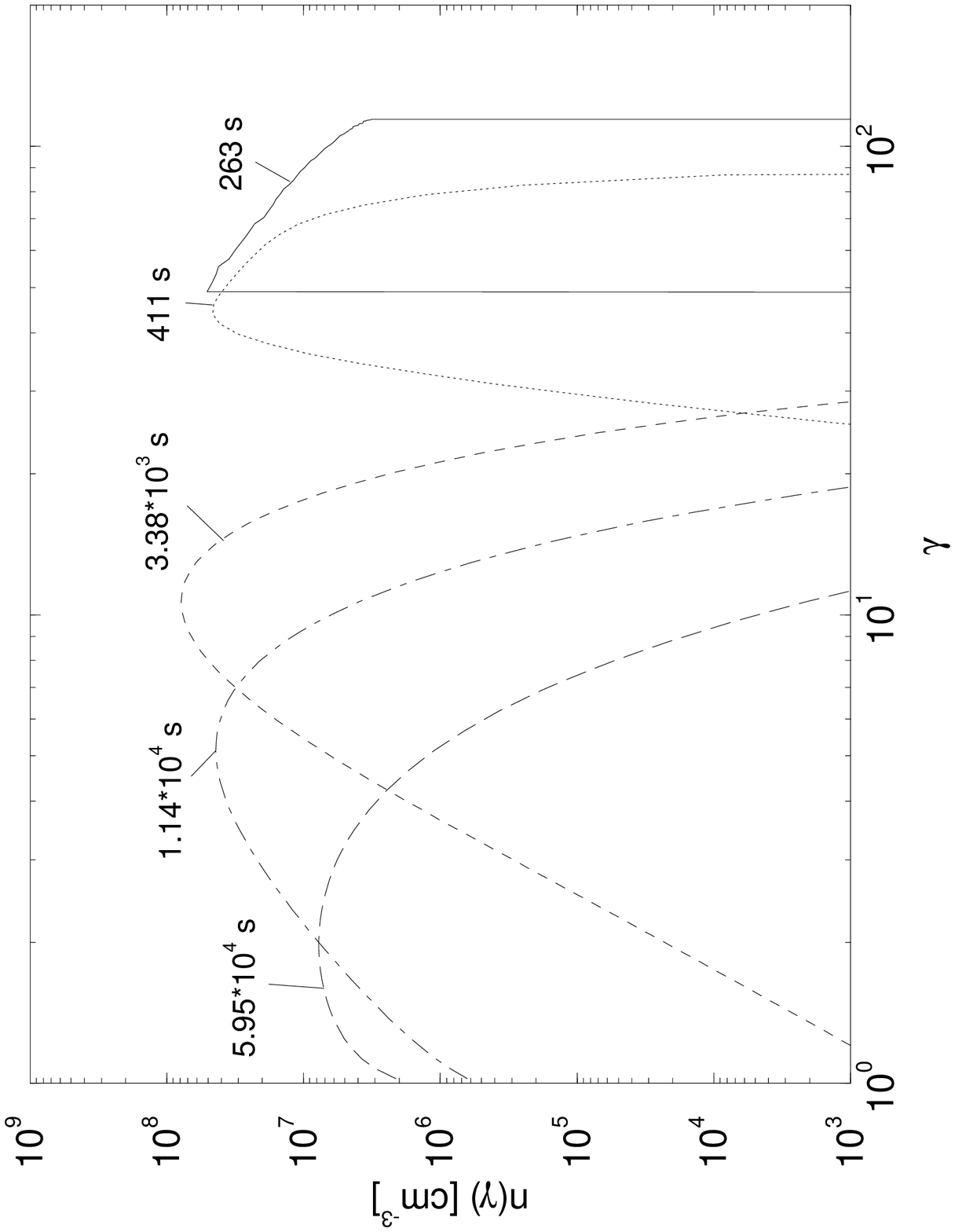} }
\caption[]{Evolution of the pair distribution functions
through the transrelativistic phase in the case of high
particle density, $n_0 = 10^9$~cm$^{-3}$. Initial conditions:
$\gepm = 200$, $\gzpm = 2 \cdot 10^4$, $s = 2$, $R_B = 5 \cdot
10^{12}$~cm, $B_0 = 1$~G, $z_i = 10^{-3}$~pc; 
$L_0 = 10^{46}$~erg~s$^{-1}$, $M = 10^8\, M_{\odot}$, $\Gamma = 10$. 
The solid line indicates the state of the plasma after the 
ultrarelativistic treatment described in BMS}
\end{figure}

\eject

\begin{figure}
\rotate[r] {
\epsfxsize=12cm
\epsffile[100 20 600 50] {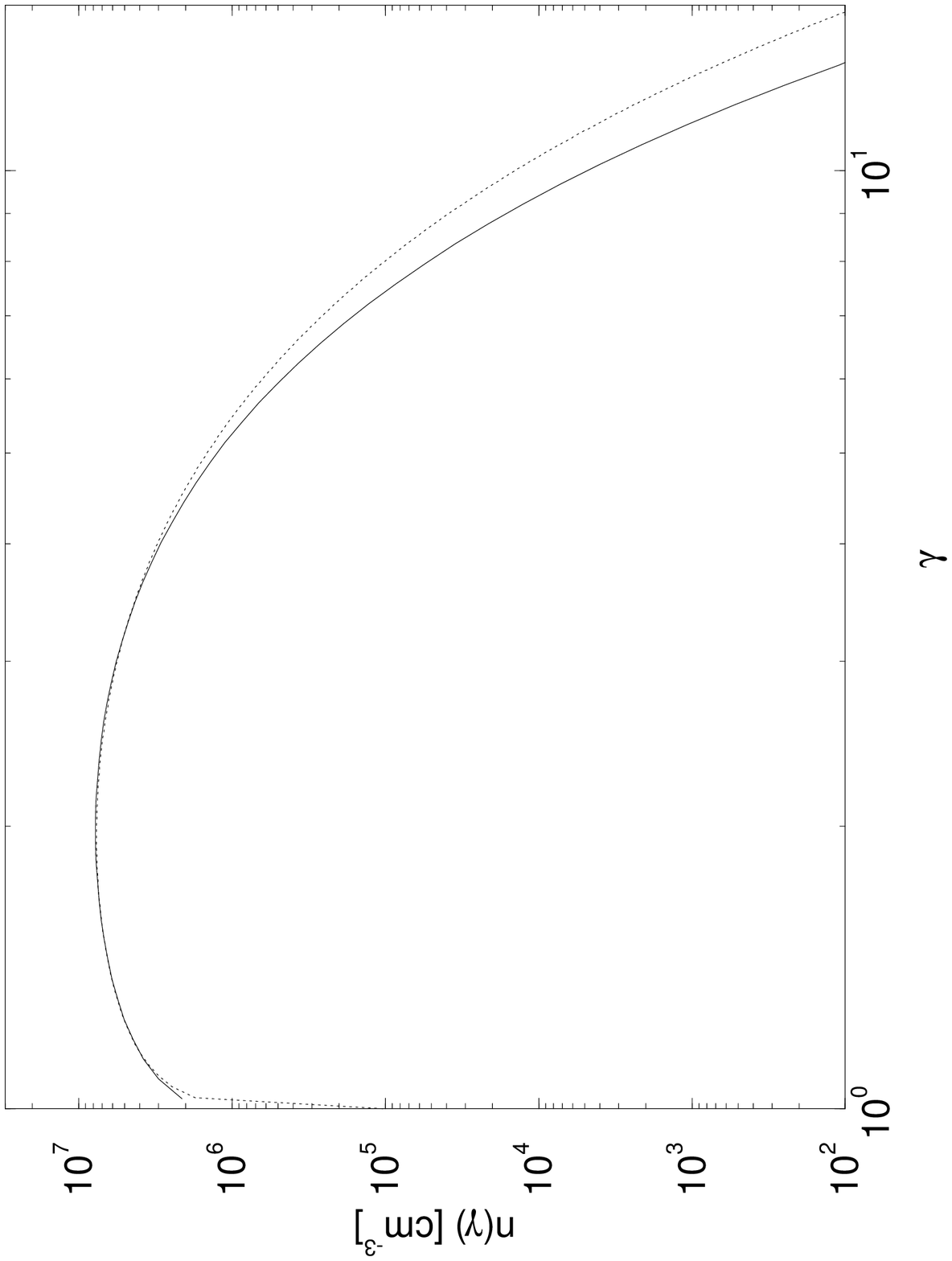} }
\caption[]{Comparison of the final state of the simulation
shown in Fig. 3 (solid) to a thermal spectrum of temperature 
$\Theta = 0.83$ (dotted)}
\end{figure}

\eject

\begin{figure}
\rotate[r] {
\epsfxsize=12cm
\epsffile[100 20 600 50] {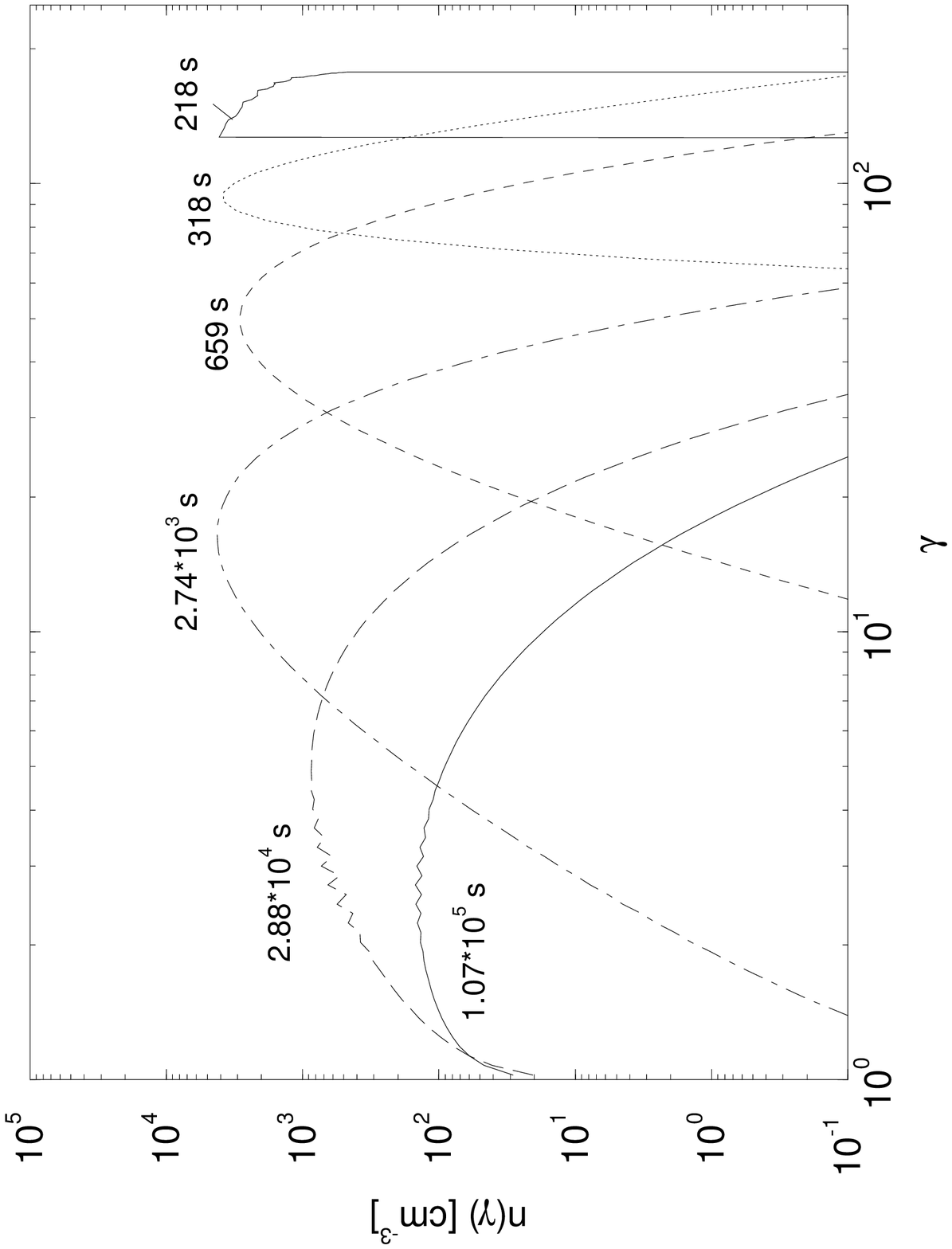} }
\caption[]{Evolution of the pair distribution functions
through the transrelativistic phase in the case of moderate
particle density, $n_0 = 10^5$~cm$^{-3}$. Initial conditions:
$\gepm = 500$, $\gzpm = 3 \cdot 10^4$, $s = 2.1$, $R_B = 2 \cdot
10^{15}$~cm, $B_0 = 1$~G, $z_i = 10^{-3}$~pc; 
$L_0 = 10^{46}$~erg~s$^{-1}$, $M = 10^8\, M_{\odot}$, $\Gamma = 10$. 
The solid line indicates the state of the plasma after the 
ultrarelativistic treatment described in BMS}
\end{figure}

\eject

\begin{figure}
\rotate[r] {
\epsfxsize=12cm
\epsffile[100 20 600 50] {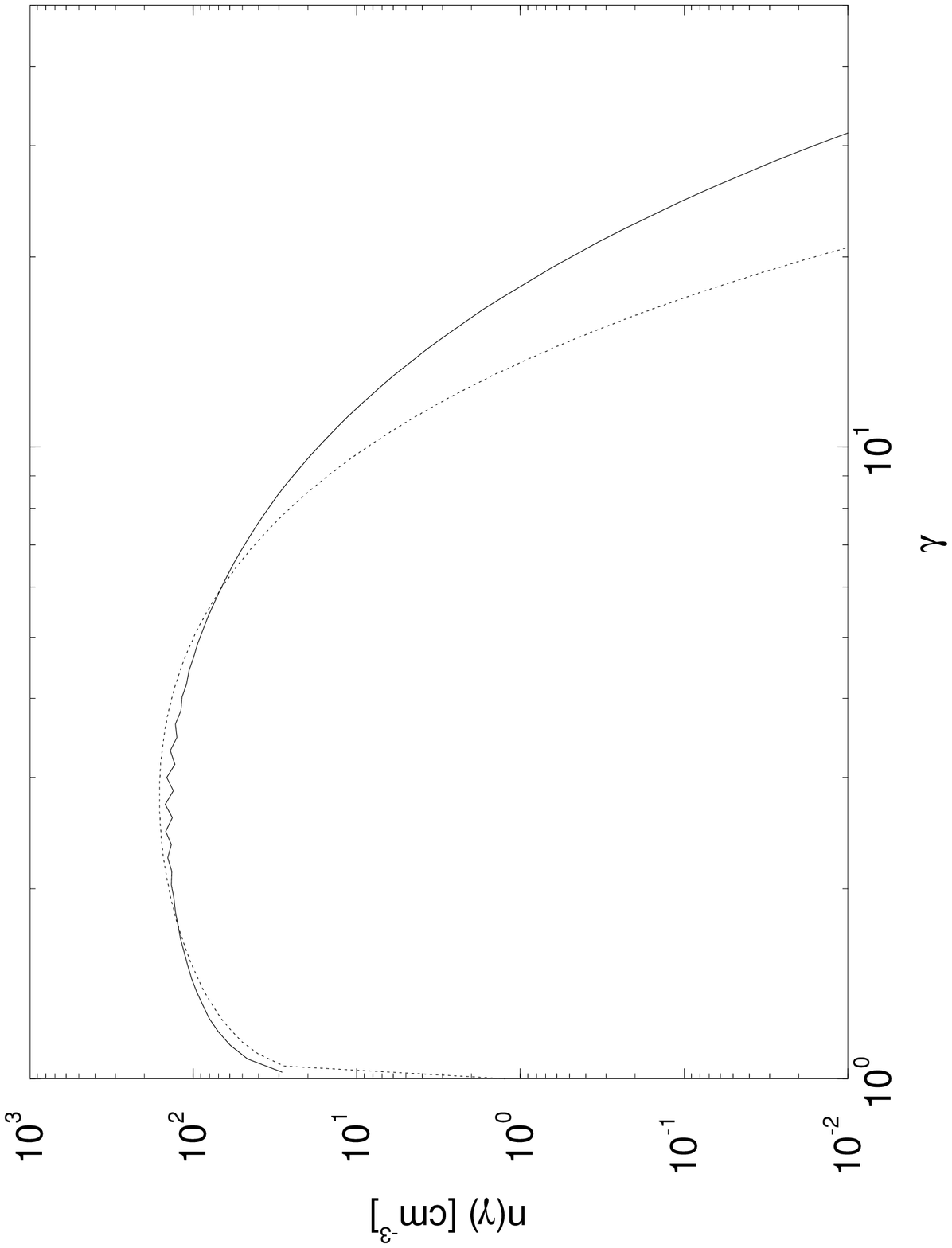} }
\caption[]{Comparison of the final state of the simulation
shown in Fig. 5 (solid) to a thermal spectrum of temperature 
$\Theta = 1.3$ (dotted)}
\end{figure}

\eject

\begin{figure}
\rotate[r] {
\epsfxsize=12cm
\epsffile[100 20 600 50] {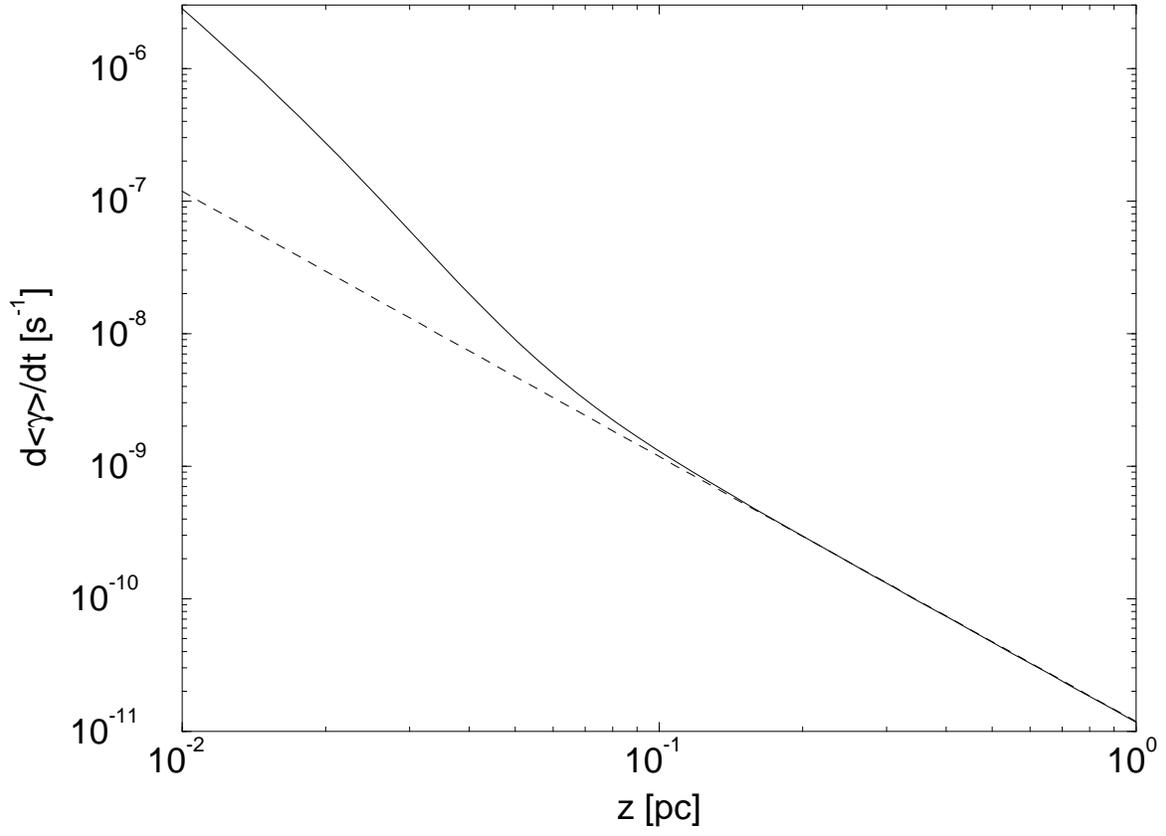} }
\caption[]{Cooling rates of a thermal pair plasma due to inverse-Compton
scattering of accretion disk photons as a function of distance from the
disk. Solid: Extended source; dashed: point source approximation. 
$L = 10^{46}$~erg~s$^{-1}$, $M = 10^8 \, M_{\odot}$, $\Theta = 2$}
\end{figure}

\eject

\begin{figure}
\rotate[r] {
\epsfxsize=12cm
\epsffile[100 20 600 50] {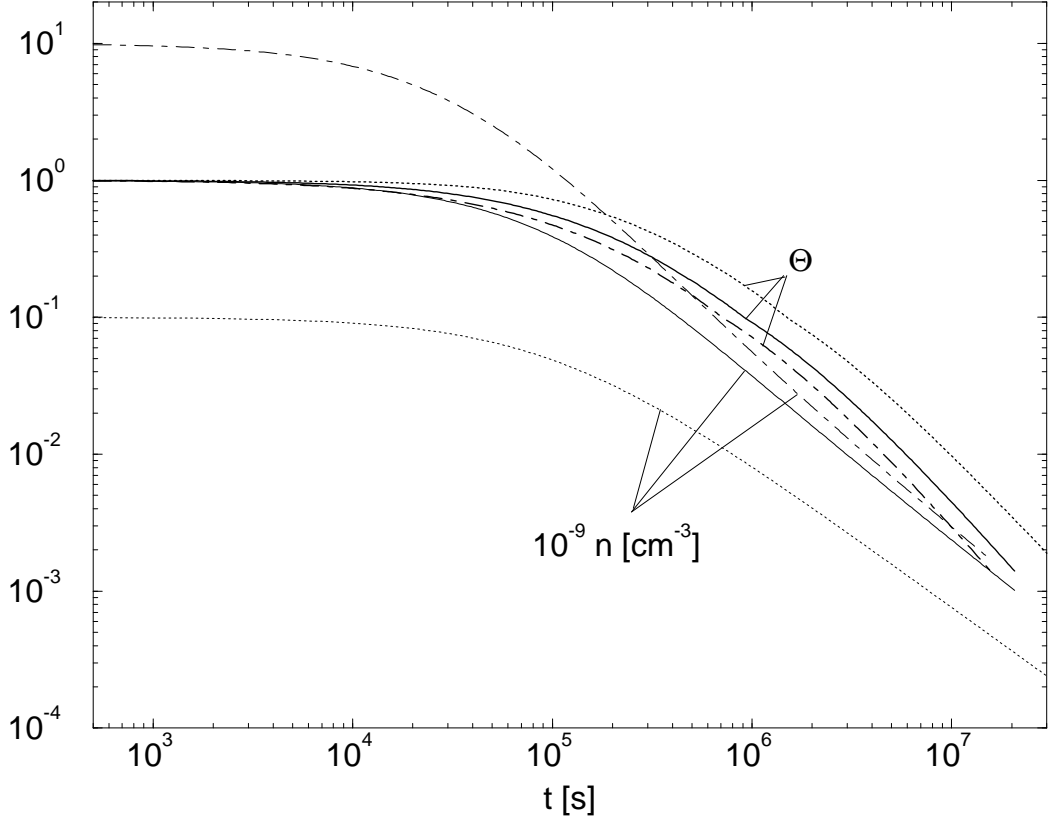} }
\caption[]{Evolution of thermal pair plasmas in a relativistic
jet. Initial temperature $\Theta_0 = 1$; $B_0 = 0.1$~G, $z_0 =
10^{-2}$~pc, $L_0 = 10^{46}$~erg~s$^{-1}$, $\Gamma = 10$, 
$\delta B / B_0 = 0.1$, $B(z) \propto z^{-1}$; initial densities 
$n_0 = 10^8$~cm$^{-3}$ (dotted), $10^9$~cm$^{-3}$ (solid), and 
$10^{10}$~cm$^{-3}$ (dot-dashed), respectively. Thick curves
show the temperature evolution, thin curves the density (multiplied
by $10^{-9}$)} 
\end{figure}

\eject

\begin{figure}
\rotate[r] {
\epsfxsize=12cm
\epsffile[100 20 600 50] {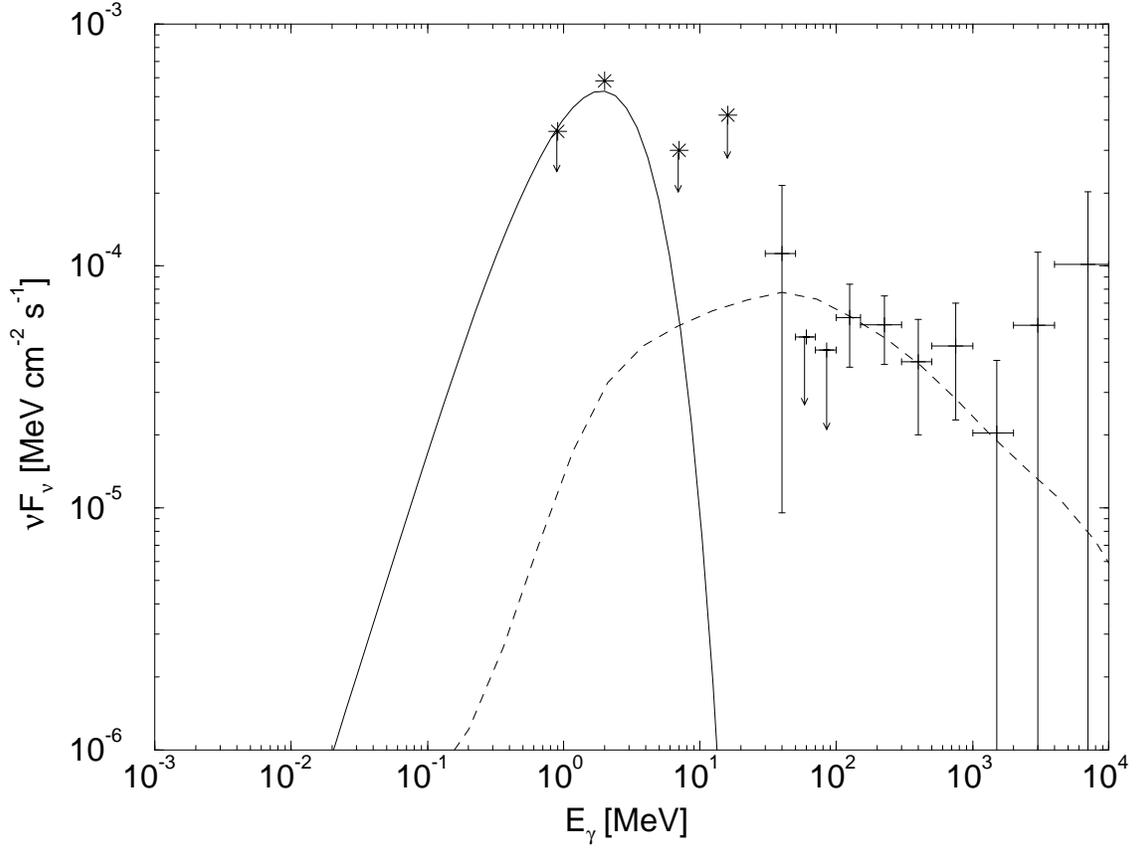} }
\caption[]{The COMPTEL and EGRET spectrum of PKS~0208-512 during 
Phase II. Dashed: high-energy (predominantly external
inverse-Compton) spectrum from ultrarelativistic plasma blobs; 
solid: inverse-Compton spectrum from the later stages of the
jet evolution, under the assumption of $\sim 10$~\% of the jet 
being filled with transrelativistic pair plasma}
\end{figure}

\end{document}